%% file: main.tex
\documentclass[11pt]{article}

\usepackage[utf8]{inputenc}
\usepackage{microtype}
\usepackage{graphicx}
\usepackage{subcaption}
\usepackage{booktabs}
\usepackage{microsoft-tech-report}
\usepackage{hyperref}
\usepackage{url}
\usepackage{amsmath}
\usepackage{amssymb}
\usepackage{amsfonts}
\usepackage{nicefrac}
\usepackage{xspace}
\usepackage{pifont}
\usepackage{multirow}
\usepackage{float}
\usepackage{enumitem}

\hypersetup{
  colorlinks=true,
  linkcolor=msftblue,
  citecolor=msftblue,
  urlcolor=msftblue,
  pdftitle={Resource2Skill: Distilling Executable Skills from Human-Created Resources for Software Agents},
  pdfauthor={Yijia Fan, Zonglin Di, Zimo Wen, Yifan Yang, Mingxi Cheng, Qi Dai, Bei Liu, Kai Qiu, Yue Dong, Ji Li, Chong Luo}
}

\newcommand{\ourmethod}{\textsc{Resource2Skill}\xspace}
\newcommand{\skilllib}{\Sigma_{\mathcal{D}}}

\newcommand{\cmark}{\ding{51}}

\newcommand{\dpos}[1]{\textcolor{green!55!black}{#1}}

\newcommand{\casegraphic}[1]{%
  \includegraphics[width=\linewidth,height=0.36\textheight,keepaspectratio]{#1}%
}
\newcommand{\webcasegraphic}[1]{%
  \includegraphics[width=\linewidth,height=0.58\textheight,keepaspectratio,trim=0 1900 0 0,clip]{#1}%
}

\techreportshorttitle{Resource2Skill}

\begin{document}
\thispagestyle{empty}

\noindent
\begin{minipage}[c]{0.5\linewidth}
\raggedright
\raisebox{-0.5\height}{\msftbrandmark}
\end{minipage}
\begin{minipage}[c]{0.49\linewidth}
\raggedleft
{\msftdatefont\small\color{msftgray}May 2026}
\end{minipage}\par
\vspace{0.35em}
\noindent{\color{msftline}\rule{\linewidth}{0.8pt}\par}

\vspace{1.0em}
\begin{center}
{{\msfttitlefont\fontsize{21}{25}\selectfont\color{msftdark}
Resource2Skill: Distilling Executable Skills\\
from Human-Created Resources\\
for Software Agents\par}}
\vspace{1.25em}

{\normalsize\rmfamily\color{msftdark}
Yijia Fan$^{1,*}$ \hspace{0.75em}
Zonglin Di$^{2,*}$ \hspace{0.75em}
Zimo Wen$^{3,*}$ \hspace{0.75em}
Yifan Yang$^{1,+}$\\[-0.1em]
Mingxi Cheng$^{1}$ \hspace{0.75em}
Qi Dai$^{1}$ \hspace{0.75em}
Bei Liu$^{1}$ \hspace{0.75em}
Kai Qiu$^{1}$\\[-0.1em]
Yue Dong$^{1}$ \hspace{0.75em}
Ji Li$^{1}$ \hspace{0.75em}
Chong Luo$^{1}$\par
}
\vspace{0.22cm}

{\footnotesize\rmfamily\color{msftgray}
$^{1}$ Microsoft Research \quad
$^{2}$ University of California, Santa Cruz \quad
$^{3}$ Shanghai Jiao Tong University\quad
}
\end{center}

\vspace{0.45em}
\begin{msfttitlebox}
\setlength{\parindent}{0cm}
\setlength{\parskip}{0.14cm}
\raggedright
\nohyphens

\input{abstract.tex}

\vspace{0.14cm}
{\setlength{\parskip}{0.06cm}\small
{\msftmetalabel{Code}\url{https://aka.ms/Resource2Skill}\par}
{\msftmetalabel{Correspondence}\href{mailto:yifanyang@microsoft.com}{yifanyang@microsoft.com}\par}
}
\vspace{0.08cm}
{\footnotesize\rmfamily\itshape\color{msftgray}
$^*$ Equal contribution. \quad
$^+$ Corresponding author.\par
}
\end{msfttitlebox}

\begin{figure}[t]
  \centering
  \includegraphics[width=\textwidth]{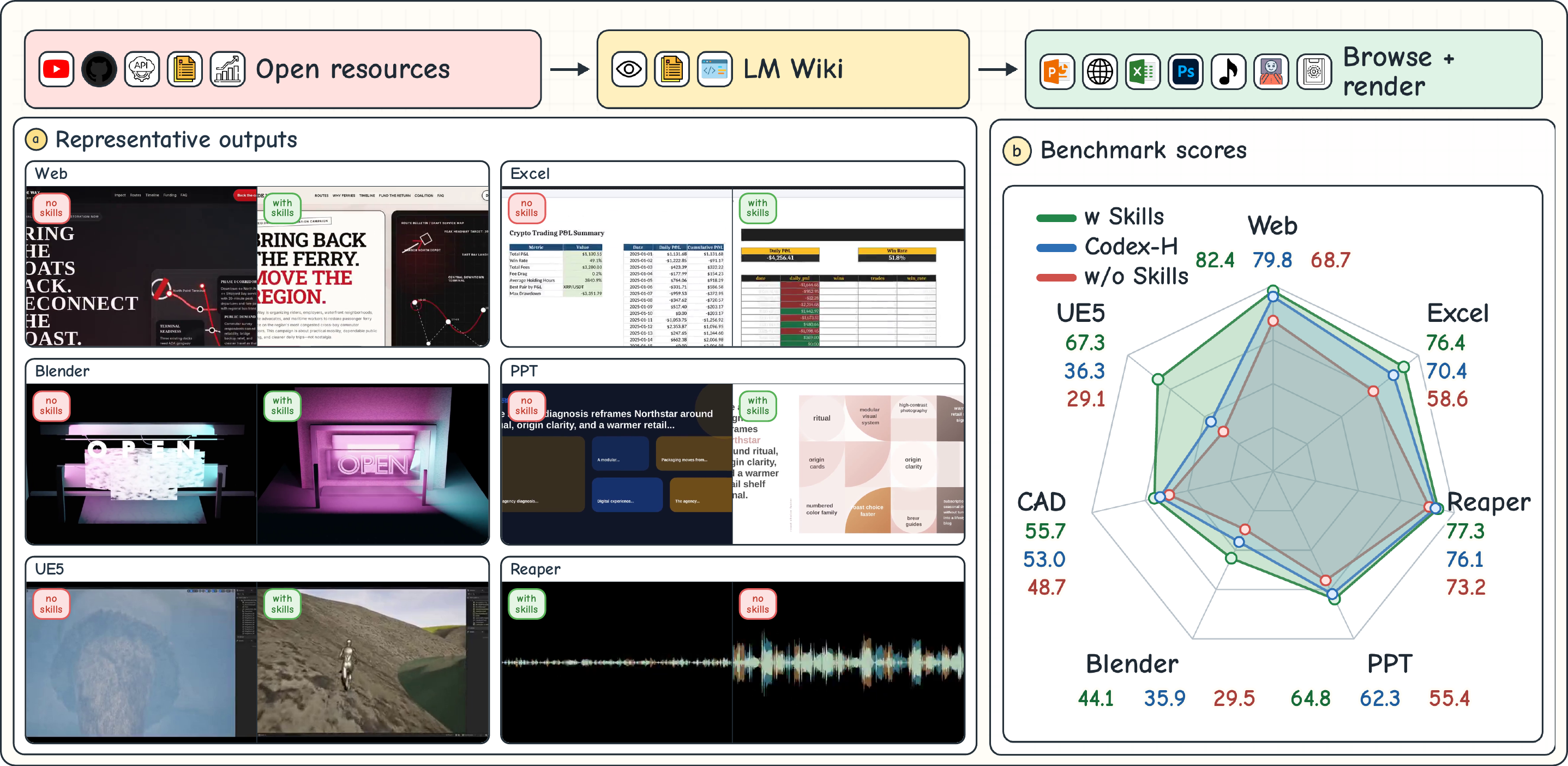}
  \caption{\textbf{\ourmethod{}} distills multimodal resources into a hierarchical Skill Wiki across seven creative software domains.}
  \label{fig:teaser}
\end{figure}

\input{intro.tex}

\section{Related Work}

\textbf{LM agents and reusable memory.} Recent agents combine reasoning with tool invocation, executable actions, and execution feedback \citep{yao2023react,karpas2022mrkl,schick2023toolformer,patil2023gorilla,qin2023toolllm,shen2023hugginggpt,gao2023pal,liang2023codeaspolicies,ahn2022saycan,shinn2023reflexion,madaan2023selfrefine}, and are now evaluated on realistic web and software-engineering tasks \citep{yao2022webshop,deng2023mind2web,zhou2024webarena,liu2023agentbench,jimenez2024swebench,yang2024sweagent}. Procedural knowledge in these systems is largely implicit in weights, prompts, or interaction history; retrieval and memory work \citep{robertson2009probabilistic,guu2020realm,lewis2020rag,borgeaud2022retro,karpukhin2020dpr,park2023generativeagents,packer2023memgpt} and skill-library agents \citep{wang2023voyager,gao2023pal,liang2023codeaspolicies} make it explicit, but typically rely on self-generated traces rather than external human references.

\textbf{Procedural knowledge from human-created resources.} Instructional videos \citep{miech2019howto100m,zhukov2019crosstask,tang2019coin,grauman2022ego4d,fan2022minedojo}, web pages \citep{nakano2021webgpt}, and public code with documentation \citep{chen2021codex,husain2019codesearchnet} carry rich procedural supervision, but their raw form---long videos, descriptive articles, scaffolding-heavy repositories, outcome-only artifacts---is poorly matched to agent execution and is usually consumed as pretraining or retrieval context. \ourmethod instead distills them into a validated multimodal wiki, turning source type, modality, library scale, and selection strategy into controlled design variables.

\textbf{Comparison with skill-library frameworks.} \textsc{Voyager}~\citep{wang2023voyager}, \textsc{AWM}~\citep{wang2024awm}, \textsc{ASI}~\citep{wang2025asi}, and \textsc{SkillFlow}~\citep{zhang2026skillflow} grow text- or code-only libraries online from a single domain's agent traces or failures and retrieve by dense similarity or repair rules; Anthropic \emph{Agent Skills}~\citep{anthropic2025skills} ships hand-authored text/code/asset bundles without automatic acquisition; \textsc{SkillFoundry}~\citep{shen2026skillfoundry}, closest in spirit, mines text/code skills offline into a top-down knowledge tree for scientific computing. Our work differs by combining offline-mined multimodal skills (videos, repositories, articles, reference artifacts), a hierarchical wiki interface with hierarchy-then-LM selection, artifact-level evaluation by vision and audio judges, and controlled online gap-filling, across seven authoring domains.

\section{Method}
\label{sec:method}

\begin{figure}[t]
\centering
\includegraphics[width=\textwidth]{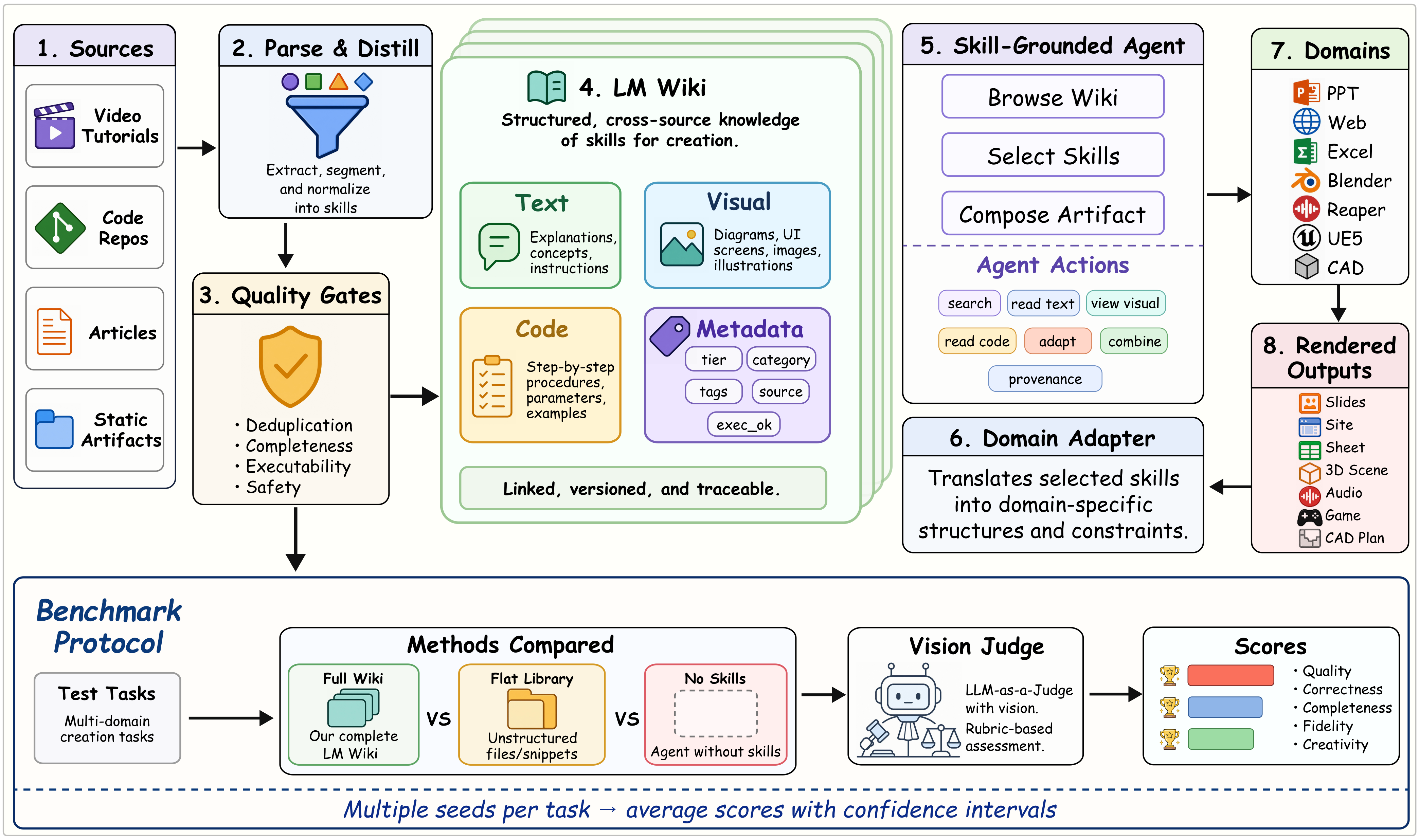}
\caption{\textbf{\ourmethod{} pipeline.} A construction operator $(f_\theta, A_\mathcal{D})$ distills resources into the hierarchical Skill Wiki; \textsc{MetaBrowse} retrieves candidates and the language model selects from text/visual/code views, applied through MCP to a domain backend. The same operator is reused online when the offline pool is insufficient.}
\label{fig:pipeline}
\end{figure}

We instantiate \ourmethod as four stages (Figure~\ref{fig:pipeline}): construction, wiki organization, selection, and execution. The same construction operator is reused online when the offline pool is insufficient, so online acquisition adds no separate pipeline, and all four stages share a single MCP-mediated browse-select-execute interface over domain-specific backends.

\subsection{Hierarchical Multimodal Skill Wiki}
\label{sec:lm-wiki}

A \emph{skill} is a tuple
\[
s \;=\; \bigl(p,\; x_{\text{text}},\; x_{\text{visual}},\; x_{\text{code}},\; m\bigr),
\]
where $p$ is the path of $s$ in a domain-specific taxonomy $\mathcal{T}_{\mathcal{D}}$ and $m$ is metadata used for filtering, auditing, and provenance. The three content views are complementary: $x_{\text{text}}$ states name, mechanism, applicability, inputs, and expected effects; $x_{\text{visual}}$ provides thumbnails, screenshots, rendered previews, or diagrams; $x_{\text{code}}$ contains executable or adaptable procedure fragments (visual and code fields may be empty for reference-only entries).

The taxonomy is domain-specific (e.g., PPT organizes by layout, typography, and motion; Blender by geometry, material, lighting, and composition) but the browse-and-read interface is shared. The full library for domain $\mathcal{D}$ is
\[
\skilllib \;=\; \bigl\{\, s : s \text{ accepted by the construction operator}\,\bigr\},
\]
and we treat its size as an experimental variable. The on-disk realization is given in Appendix~\ref{app:schema}.

\subsection{Resource-to-Skill Construction}
\label{sec:multi-source}

A construction operator distills multimodal resources into wiki entries. The resource pool $\mathcal{R}_{\mathcal{D}}$ for domain $\mathcal{D}$ is drawn from four families: tutorial videos, source repositories, articles, and reference artifacts.

A multimodal distiller maps each resource $r \in \mathcal{R}_{\mathcal{D}}$ to candidate skills,
\[
\tilde{s}_{1:k} \;=\; f_{\theta}(r,\, \mathcal{D}),
\]
each expressed in the wiki schema above. Concretely, $f_\theta$ retrieves resources against domain-specific queries, extracts modality-specific evidence (key frames, code regions and parameter signatures, prose passages, rendered exemplars), distills it into $(p, x_{\text{text}}, x_{\text{visual}}, x_{\text{code}}, m)$ via a vision-capable LM, and normalizes the result. A domain-specific predicate $A_{\mathcal{D}}$ then enforces five checks---completeness, traceable provenance, deduplication, modality consistency, and structural executability of the code field when present:
\[
\skilllib \;=\; \Bigl\{\, s = \mathrm{normalize}(\tilde{s}) \;:\; \tilde{s} = f_{\theta}(r, \mathcal{D}),\; r \in \mathcal{R}_{\mathcal{D}},\; A_{\mathcal{D}}(\tilde{s}) = 1 \,\Bigr\}.
\]
Per-domain library sizes and the executability protocol are summarized in Appendix~\ref{app:domains}. The same operator $(f_\theta, A_\mathcal{D})$ is reused at test time for online acquisition, so source mix becomes a controlled design variable rather than a hidden implementation detail.

\subsection{Selecting and Composing Skills}
\label{sec:metabrowse}

Given a brief $q$, the agent must choose a small subset of wiki entries to compose. \textsc{MetaBrowse} uses the wiki's hierarchical organization in two stages: a lexical scorer narrows the candidate set to a topically relevant region of the taxonomy, then a language model selects a subset to compose. The first-stage score combines the entry's name, tags, applicability text, and---critically---its taxonomy path $p(s)$,
\[
\mathcal{C}_K(q) \;=\; \operatorname{TopK}_{s \in \skilllib} \mathrm{BM25}\!\bigl(q,\; \mathrm{name}(s) \oplus \mathrm{tags}(s) \oplus \mathrm{applicability}(s) \oplus p(s)\bigr),
\]
so the wiki tree directly favours skills sitting in topically relevant subtrees rather than treating the library as a flat list. The language model then reads structured evidence for candidates in $\mathcal{C}_K(q)$ and selects a subset to compose,
\[
\mathcal{S}(q) \;=\; \pi_{\phi}\!\bigl(q,\; \{\Phi(s) : s \in \mathcal{C}_K(q)\}\bigr),
\]
where $\Phi(s)$ contains metadata and whichever text, visual, and code views are exposed by the current configuration. Selection is a subset rather than a ranking: the language model can pick zero skills if no candidate is a good fit.

\subsection{Execution and Online Acquisition}
\label{sec:executor}
\label{sec:online-acquisition}

The agent and domain adapter share one MCP tool surface. The wiki side exposes a small set of discovery actions (list categories, list skills with metadata cards, read per-modality content) together with a search action wrapping the BM25 shortlist; the domain side exposes a single \emph{apply} action backed by a per-domain capabilities manifest, with structured not-applicable returns for missing capabilities. A run follows the same control loop in every domain (plan, \textsc{MetaBrowse}, apply, render), and selected skill code, when present, executes directly against the live MCP server with no language-model translation between selection and execution. Reference-only skills remain useful: the agent adapts their text and visual evidence to write its own code, and provenance is retained through the selected skill identifiers.

When $\mathcal{C}_K(q)$ contains no adequate candidate, the same operator $(f_\theta, A_\mathcal{D})$ is invoked online: targeted queries are issued on the same resource families, returned resources are distilled into temporary candidates, validated by $A_\mathcal{D}$, and exposed as a separate online pool for the current task or evaluation split. Offline and online pools are kept separate throughout the experiments, so online acquisition is a controlled gap-filler on capability regions known to be insufficient, rather than uncontrolled context expansion at test time.

\section{Experiments}
\label{sec:experiments}

We evaluate \ourmethod across seven authoring domains and four agent backends, comparing it against a no-skill ablation and two off-the-shelf agentic harnesses, and isolating the wiki interface, source mix, entry format, library scale, online acquisition, and selection strategy in controlled ablations.

\subsection{Setup}
\label{sec:setup}

\paragraph{Domains and briefs.} We evaluate on seven authoring domains: slide design (PPT), 2D drafting (CAD), web (HTML/CSS/JS), spreadsheet authoring (Excel), 3D scenes (Blender), real-time 3D (UE5), and audio production (Reaper). Per-domain backends and rendering paths are in Appendix~\ref{app:domains}. Each domain has a screened pool of $80$ task briefs with no overlap with the resource corpora used to build the library; the brief author is blind to the wiki and the agent. The main comparison and the scaling and online/offline studies use a matched $N{=}80$ subset per domain; ablations use matched $N{=}40$ subsets. All conditions in a comparison share brief IDs, so within-table deltas are paired by construction.

\paragraph{Compared systems.} We sweep four agent backends---GPT-5.5, GPT-5.4, GPT-5.4 Mini, and GPT-5.4 Nano---against four systems. \textsc{w Skills} is the full \ourmethod pipeline. \textsc{w/o Skills} is the same agent solving tasks through free-form code over the domain apply tool, with no skill library. \textsc{ClaudeCode-H} and \textsc{Codex-H} are the off-the-shelf Claude Code and Codex agentic harnesses (the \emph{-H} suffix denotes \emph{harness}, not human rater); CLI versions, invocation contract, and matched configuration are in Appendix~\ref{app:harness}. All agent and judge calls use temperature $0$ and reasoning effort \emph{low}. Studies and ablations use GPT-5.4 unless stated otherwise.

\paragraph{Judge and aggregation.} Non-audio artifacts are judged by a GPT-5.4 vision judge; Reaper is judged by an audio-capable GPT-4o-series judge. The judge is blinded to the system label and sees only the brief and the rendered artifact. Each domain has its own five-axis rubric (full protocol and human--judge agreement in Appendices~\ref{app:metrics} and~\ref{app:judge}); rubric scores in $0$--$10$ are reported as percentages, and the \textbf{overall} score is the unweighted arithmetic mean of the five axes. A run that fails to produce a scorable artifact---no artifact returned, or an artifact below a per-domain minimum-quality threshold---is treated as a \emph{failure} and folds in at $\text{overall}{=}0$.

\subsection{Main Comparison: With-Skill vs Without-Skill}
\label{sec:exp-skill}
\label{sec:exp-main}  

The first question is whether the skill library contributes measurable lift. We compare the four systems (\textsc{w Skills}, \textsc{w/o Skills}, \textsc{ClaudeCode-H}, \textsc{Codex-H}) on the matched-brief suite of $N{=}80$ tasks per domain, repeated across the four agent backends. The judge, brief set, and decoding seed are held fixed within each model-domain cell, so score differences isolate the effect of the execution interface and skill access. Table~\ref{tab:main-final-v2} reports the overall score as a percentage.

\begin{table}[t]
\centering
\caption{\textbf{Main comparison, overall score (\%).} \emph{Avg.}\ is the unweighted mean over all seven domain columns. \textbf{Bold} marks the best system per column within each backend group. Per-cell paired outcome counts and Wilcoxon $p$-values are tabulated in Appendix~\ref{app:per-cell-stats}.}
\label{tab:main-final-v2}
\small
\resizebox{\textwidth}{!}{
\begin{tabular}{ll|ccccc|cc|c}
\toprule
\textbf{Model} & \textbf{System}
& \textbf{Web}
& \textbf{Excel}
& \textbf{Reaper}
& \textbf{PPT}
& \textbf{Blender}
& \textbf{CAD}
& \textbf{UE5}
& \textbf{Avg.} \\
\midrule
\multirow{4}{*}{GPT-5.5}
& \textsc{w Skills}     & $82.8$          & $\mathbf{61.3}$ & $\mathbf{77.6}$ & $\mathbf{67.5}$ & $\mathbf{53.1}$ & $\mathbf{48.7}$ & $\mathbf{69.5}$ & $\mathbf{65.8}$ \\
& \textsc{w/o Skills}   & $69.4$          & $58.2$          & $73.1$          & $53.9$          & $35.6$          & $42.6$          & $30.2$          & $51.9$ \\
& \textsc{ClaudeCode-H} & $\mathbf{83.5}$ & $59.8$          & $76.2$          & $63.4$          & $45.7$          & $47.1$          & $35.9$          & $58.8$ \\
& \textsc{Codex-H}      & $81.6$          & $60.5$          & $76.9$          & $64.1$          & $44.9$          & $47.0$          & $36.0$          & $58.7$ \\
\midrule
\multirow{4}{*}{GPT-5.4}
& \textsc{w Skills}     & $\mathbf{82.4}$ & $\mathbf{76.4}$ & $\mathbf{77.3}$ & $\mathbf{64.8}$ & $\mathbf{44.1}$ & $\mathbf{55.7}$ & $\mathbf{67.3}$ & $\mathbf{66.9}$ \\
& \textsc{w/o Skills}   & $68.7$          & $58.6$          & $73.2$          & $55.4$          & $29.5$          & $48.7$          & $29.1$          & $51.9$ \\
& \textsc{ClaudeCode-H} & $81.6$          & $69.2$          & $75.8$          & $61.6$          & $36.7$          & $53.3$          & $35.7$          & $59.1$ \\
& \textsc{Codex-H}      & $79.8$          & $70.4$          & $76.1$          & $62.3$          & $35.9$          & $53.0$          & $36.3$          & $59.1$ \\
\midrule
\multirow{4}{*}{GPT-5.4 Mini}
& \textsc{w Skills}     & $\mathbf{67.6}$ & $\mathbf{45.7}$ & $\mathbf{62.6}$ & $\mathbf{52.4}$ & $\mathbf{28.8}$ & $\mathbf{50.3}$ & $\mathbf{55.9}$ & $\mathbf{51.9}$ \\
& \textsc{w/o Skills}   & $55.2$          & $42.4$          & $58.3$          & $45.9$          & $18.7$          & $45.6$          & $23.7$          & $41.4$ \\
& \textsc{ClaudeCode-H} & $66.8$          & $44.8$          & $61.4$          & $49.3$          & $24.6$          & $48.9$          & $26.6$          & $46.1$ \\
& \textsc{Codex-H}      & $65.4$          & $45.2$          & $61.8$          & $50.8$          & $23.9$          & $49.7$          & $26.9$          & $46.2$ \\
\midrule
\multirow{4}{*}{GPT-5.4 Nano}
& \textsc{w Skills}     & $\mathbf{49.5}$ & $\mathbf{34.6}$ & $\mathbf{50.5}$ & $41.3$          & $\mathbf{15.8}$ & $\mathbf{51.4}$ & $\mathbf{56.3}$ & $\mathbf{42.8}$ \\
& \textsc{w/o Skills}   & $42.3$          & $31.8$          & $48.7$          & $38.6$          & $12.2$          & $45.1$          & $24.5$          & $34.7$ \\
& \textsc{ClaudeCode-H} & $48.2$          & $33.4$          & $49.8$          & $40.2$          & $14.5$          & $49.7$          & $26.2$          & $37.4$ \\
& \textsc{Codex-H}      & $47.6$          & $34.1$          & $50.3$          & $\mathbf{42.1}$ & $13.9$          & $50.3$          & $26.9$          & $37.9$ \\
\bottomrule
\end{tabular}
}
\end{table}

\paragraph{Skills dominate across backends.} \textsc{w Skills} beats \textsc{w/o Skills} in all $28$ main-aggregate model--domain cells, averaging $56.8\%$ versus $45.0\%$, a $+11.9$-point lift. Both off-the-shelf harnesses raise the no-skill agent on their own but remain consistently below \textsc{w Skills}: \textsc{Codex-H} reaches $50.5\%$ and \textsc{ClaudeCode-H} $50.4\%$ on the same matched briefs, and \textsc{w Skills} still beats the stronger of the two in $26$ of $28$ cells. The two exceptions (GPT-5.5 Web vs.\ \textsc{ClaudeCode-H}, GPT-5.4 Nano PPT vs.\ \textsc{Codex-H}) are within one point. The dominant lift therefore comes from the curated wiki, not from the execution harness alone. In the reported paired cells, deltas of \textsc{w Skills} over \textsc{w/o Skills} are statistically significant (paired Wilcoxon $p < 10^{-3}$; Appendix~\ref{app:per-cell-stats}).

\paragraph{Lift concentrates where conventions matter.} Per-domain gains are largest where authoring conventions are dense and expensive to re-derive from a prompt---Blender, Web, and UE5 at the larger backends---and largest of all on UE5 ($+30$ to $+40$ pp), where the free-form code agent rarely assembles a minimum-viable scene through the UE5 Python API and frequently returns artifacts below the per-domain minimum-quality threshold. Reaper sees the smallest gains, reflecting a relatively competent no-skill prior over the medium. A blinded human A/B study with five raters per pair on $40$ matched pairs across all seven domains ($200$ paired ratings, Appendix~\ref{app:userstudy}) corroborates the preference direction across all seven domains, with \textsc{w Skills} winning $85.5\%$ of non-tied individual ratings against \textsc{w/o Skills}. Per-axis traces of the lift are in Table~\ref{tab:main-perdim} (Appendix~\ref{app:perdim}).

\subsection{Studies}

\subsubsection{Skill-Library Scaling}
\label{sec:exp-scaling}

This study asks how performance moves as we vary the size of the skill pool. Holding the agent, judge, brief set, and wiki interface fixed, we expose progressively larger category-balanced pools from $0$ skills to the complete library and evaluate on the matched $N{=}80$ brief suite in each domain. The $0$-skill condition is the same free-form baseline as the main comparison; the final tick in panel~a denotes the full pool used by the main system.

\begin{figure}[t]
\centering
\begin{minipage}[t]{0.49\textwidth}
\centering
\includegraphics[width=\linewidth]{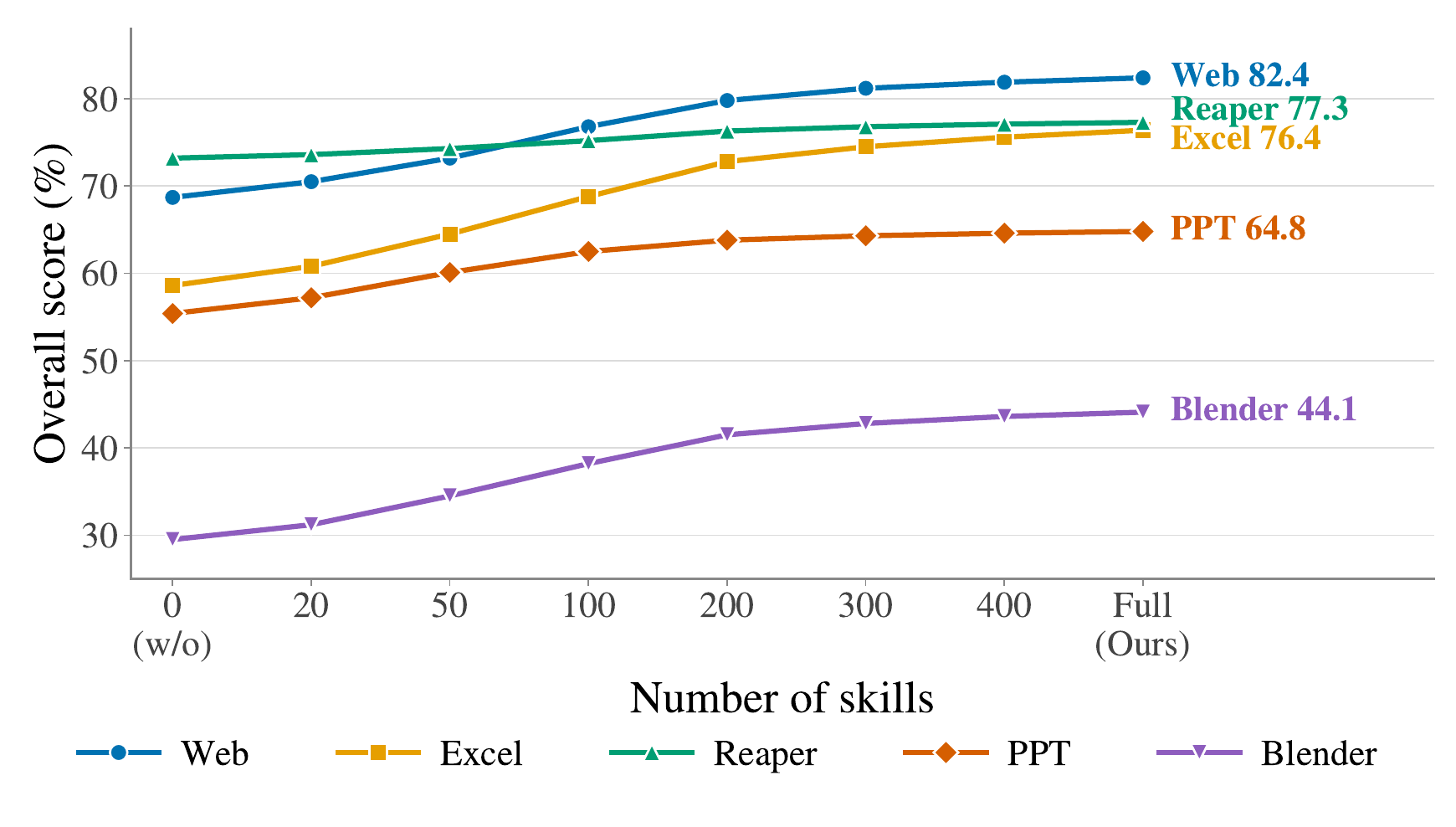}
\vspace{0.7em}
\parbox{\linewidth}{\centering\footnotesize\textbf{a, Skill-pool scaling.} The final tick, \textsc{Full}, is the complete skill pool used by \ourmethod; the table lists plotted percentages.}
\end{minipage}\hfill
\begin{minipage}[t]{0.49\textwidth}
\centering
\includegraphics[width=\linewidth]{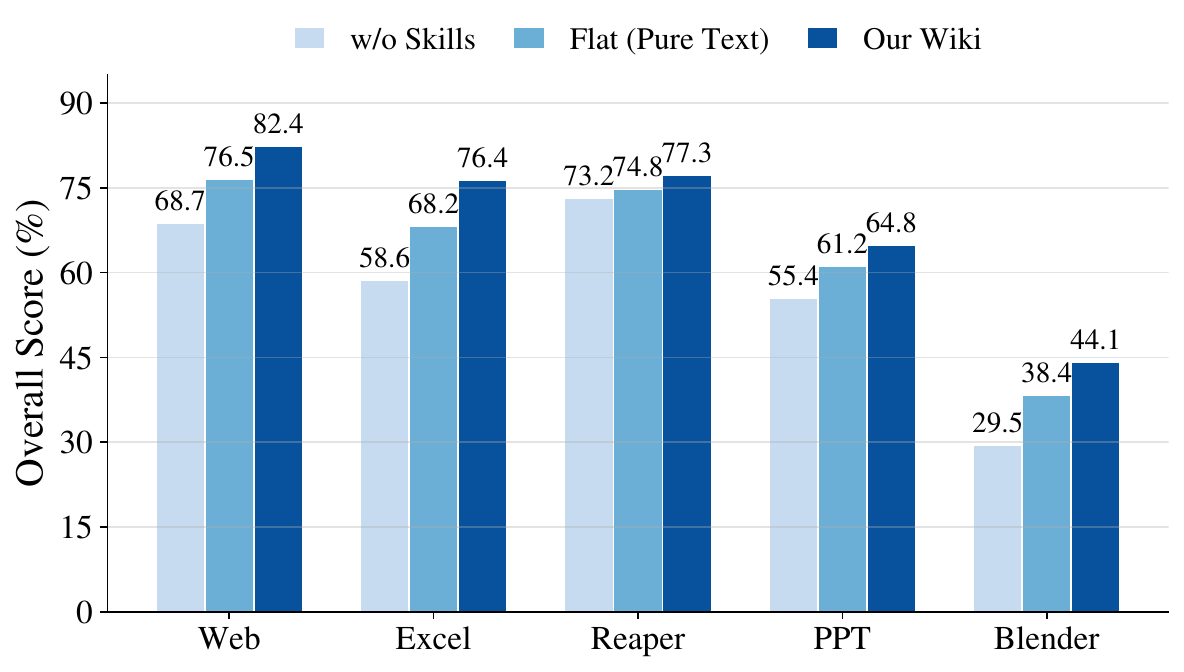}
\vspace{0.7em}
\parbox{\linewidth}{\centering\footnotesize\textbf{b, Wiki organization.} The full wiki interface is compared with no skills and flat pure-text skill access; the table lists plotted percentages.}
\end{minipage}
\end{figure}

\paragraph{Result.} Performance rises monotonically with library size in every domain and saturates near $200$ skills. The first $0\rightarrow 200$ slice carries the largest gains (between $+3.1$ pp on Reaper and $+14.2$ pp on Excel); the curve flattens after $200$ and the $400\rightarrow$\textsc{Full} step adds at most $+0.8$ pp per domain. Early entries cover common operations and recovery routines; later entries fill domain-specific gaps.

\subsubsection{Offline and Online Acquisition}
\label{sec:exp-online}

We compare \textsc{Offline-only}, which retrieves from a fixed reduced cross-domain whitelist of $891$ distilled skills, against \textsc{Offline+Online}, which adds up to $100$ skills distilled online when the offline pool is inadequate. We evaluate on two task sets: $T_{\text{standard}}$ is the regular matched benchmark; $T_{\text{novel}}$ is a stress-test suite targeting capabilities the offline pool failed to cover during preliminary screening. Online candidates are not folded back into the offline pool.

\begin{table}[t]
\centering
\caption{\textbf{Offline and online skill acquisition, overall score (\%).} $T_{\text{standard}}$ is the regular benchmark; $T_{\text{novel}}$ targets capabilities missing from the offline pool. \textbf{Bold} marks the better configuration within each task set.}
\label{tab:online-offline}
\small
\begin{tabular}{l|cc|l|cc}
\toprule
\textbf{Configuration}
& \textbf{Offline Pool}
& \textbf{Online Pool}
& \textbf{Task Set}
& \textbf{Mean Overall (\%)}
& \textbf{$\Delta$ vs.\ Base (pp)} \\
\midrule
\textsc{Offline-only}   & $891$ & $0$   & $T_{\text{standard}}$ & $65.4$ & -- \\
\textsc{Offline+Online} & $891$ & $100$ & $T_{\text{standard}}$ & $\mathbf{66.1}$ & $+0.7$ \\
\midrule
\textsc{Offline-only}   & $891$ & $0$   & $T_{\text{novel}}$ & $41.2$ & -- \\
\textsc{Offline+Online} & $891$ & $100$ & $T_{\text{novel}}$ & $\mathbf{62.8}$ & $\mathbf{+21.6}$ \\
\bottomrule
\end{tabular}
\end{table}

\paragraph{Result.} The two task sets reveal a sharp asymmetry. On $T_{\text{standard}}$, online acquisition adds $+0.7$ pp---essentially noise, since the offline pool already covers most common requests. On $T_{\text{novel}}$, the same $100$ online skills lift the mean score from $41.2\%$ to $62.8\%$ ($+21.6$ pp). Online search is a gap-filler, not a booster; we therefore default to \textsc{Offline-only} for the standard benchmark.

\subsection{Ablations}

\subsubsection{Skill + Wiki vs Skill Only}
\label{sec:exp-wiki}

\paragraph{Setup.} We compare three GPT-5.4 conditions on the full matched $N{=}80$ set per domain. \textsc{w/o Skills} disables the library; \textsc{Flat} exposes pure text skill descriptions as an unstructured list, removing category browsing, metadata, visual previews, and executable code; \textsc{Our Wiki} is the full skill-enabled wiki interface used by \ourmethod. Panel~b reports the three arms.

\paragraph{Result.} The flat text library already lifts every domain over disabling skills, confirming that reusable skill descriptions help on their own. \textsc{Our Wiki} is best in every domain and beats \textsc{Flat} by $2.5$ to $8.2$ pp, with the largest gains where category structure and code grounding matter most (Excel, Web, Blender). Hierarchical browsing narrows the search space; code and visual fields supply execution grounding that pure text under-specifies.

\subsubsection{Wiki Construction}
\label{sec:exp-source}
\label{sec:exp-format}

\paragraph{Source ablation.} The top row of Table~\ref{tab:source-updated} holds \textsc{Video} out (\textsc{Code+Article+Artifact} only) to test whether the other source families can recover video's contribution; the next four rows seed with video and add zero or one supplemental source family; the bottom row combines all four. This design measures both video's irreplaceability and the marginal utility of each supplemental source.

\begin{table}[t]
\centering
\caption{\textbf{Ablation: resource-source mix, overall score (\%).} A checkmark indicates the source family is included. The top row holds \textsc{Video} out; the bottom row is the full \ourmethod source pool. Each cell averages $N{=}40$ matched briefs. \textbf{Bold} marks the best configuration per domain.}
\label{tab:source-updated}
\small
\begin{tabular}{cccc|ccccc|c}
\toprule
\textbf{Video}
& \textbf{Code}
& \textbf{Article}
& \textbf{Artifact}
& \textbf{Web}
& \textbf{Excel}
& \textbf{Reaper}
& \textbf{PPT}
& \textbf{Blender}
& \textbf{Avg.} \\
\midrule
       & \cmark & \cmark & \cmark & $71.3$ & $61.6$ & $74.2$ & $57.4$ & $32.7$ & $59.4$ \\
\cmark &        &        &        & $81.1$ & $73.7$ & $75.6$ & $62.4$ & $41.3$ & $66.8$ \\
\cmark & \cmark &        &        & $82.0$ & $75.2$ & $76.3$ & $62.9$ & $41.6$ & $67.6$ \\
\cmark &        & \cmark &        & $81.9$ & $74.4$ & $77.8$ & $63.7$ & $42.9$ & $68.1$ \\
\cmark &        &        & \cmark & $81.6$ & $74.8$ & $76.4$ & $63.5$ & $42.4$ & $67.7$ \\
\rowcolor{blue!10}
\cmark & \cmark & \cmark & \cmark & $\mathbf{82.8}$ & $\mathbf{75.8}$ & $\mathbf{78.1}$ & $\mathbf{64.2}$ & $\mathbf{43.8}$ & $\mathbf{68.9}$ \\
\bottomrule
\end{tabular}
\end{table}

\paragraph{Source result.} Video is the non-substitutable source. Holding it out drops the average from $68.9\%$ to $59.4\%$, and the video-only library still outscores the three-source no-video library by $7.4$ points. The video-removal drop concentrates where temporal operations and visual sequencing carry signal that text under-specifies (Excel $-14.2$ pp, Web $-11.5$ pp). Beyond video, no supplemental family dominates, but the all-source pool stays $0.3$ to $0.9$ pp ahead of the strongest two-source variant in every domain; diversity adds coverage insurance on top of video.

\paragraph{Matched-budget representation ablation.} This ablation isolates multimodal skill content from curated text memory. All conditions share the resource pool, accepted skill IDs, wiki frontmatter/metadata, BM25-then-LM budget, GPT-5.4 agent, and judge; only the post-retrieval content (text, visual thumbnail, executable code) changes. Thus \textsc{Text} is a strong curated text-memory baseline with matched metadata and retrieval.

\begin{table}[t]
\centering
\caption{\textbf{Matched-budget representation ablation, overall score (\%).} All rows use the same resource pool, the same accepted skill IDs, the same wiki frontmatter and metadata, the same BM25-then-LM retrieval budget, and the same agent. A checkmark indicates the modality exposed to the agent post-retrieval; \textsc{Full} is the full wiki entry. Each cell averages $N{=}40$ matched briefs.}
\label{tab:format-updated}
\small
\begin{tabular}{ccc|ccccc|c}
\toprule
\textbf{Text}
& \textbf{Visual}
& \textbf{Code}
& \textbf{Web}
& \textbf{Excel}
& \textbf{Reaper}
& \textbf{PPT}
& \textbf{Blender}
& \textbf{Avg.} \\
\midrule
\cmark &        &        & $80.3$ & $72.7$ & $73.6$ & $60.5$ & $37.8$ & $65.0$ \\
\cmark & \cmark &        & $81.6$ & $74.1$ & $74.8$ & $63.4$ & $40.7$ & $66.9$ \\
\cmark &        & \cmark & $80.9$ & $75.2$ & $75.9$ & $61.8$ & $41.4$ & $67.0$ \\
\rowcolor{blue!10}
\cmark & \cmark & \cmark & $\mathbf{82.8}$ & $\mathbf{75.8}$ & $\mathbf{78.1}$ & $\mathbf{64.2}$ & $\mathbf{43.8}$ & $\mathbf{68.9}$ \\
\bottomrule
\end{tabular}
\end{table}

\paragraph{Result.} \textsc{Text} reaches $65.0\%$ because it already inherits applicability and routing cues. Visuals add $+1.9$ pp, code adds $+2.0$ pp, and \textsc{Full} ranks first in every domain at $68.9\%$, attributing the remaining gain to multimodal skill content rather than curated memory alone.

\subsubsection{Selection Strategy}
\label{sec:exp-selection}

\paragraph{Setup.} On the same matched $N{=}40$ subset we compare six selection strategies under a fixed library, agent, judge, and candidate budget. \textsc{Ours} is the hierarchy-then-LM \textsc{MetaBrowse} selector. \textsc{BM25} and \textsc{Embed} are retrieval-only baselines (lexical and dense, the latter a standard dense-retrieval RAG over the skill library); \textsc{BM25+Embed} combines them via dense reranking on a lexical shortlist. \textsc{Random-FullPool} samples uniformly from the skill pool; \textsc{No-Skill} disables skills as a floor reference.

\begin{table}[t]
\centering
\caption{\textbf{Ablation: selection strategy, overall score (\%).} Each cell averages $N{=}40$ matched briefs under the same agent, judge, library, and candidate budget. \textbf{Bold} marks the best strategy per domain.}
\label{tab:selection-updated-aligned}
\small
\begin{tabular}{l|ccccc|c}
\toprule
\textbf{Method}
& \textbf{Excel}
& \textbf{Web}
& \textbf{PPT}
& \textbf{Blender}
& \textbf{Reaper}
& \textbf{Avg.} \\
\midrule
\rowcolor{blue!10}
\textsc{Ours}   & $\mathbf{75.8}$ & $\mathbf{82.8}$ & $\mathbf{64.2}$ & $\mathbf{43.8}$ & $\mathbf{78.1}$ & $\mathbf{68.9}$ \\
\midrule
\textsc{BM25}            & $70.8$ & $80.5$ & $60.4$ & $41.5$ & $76.8$ & $66.0$ \\
\textsc{Embed}             & $63.5$ & $75.2$ & $48.1$ & $37.8$ & $75.2$ & $60.0$ \\
\textsc{BM25+Embed}        & $69.1$ & $81.6$ & $57.5$ & $36.2$ & $76.6$ & $64.2$ \\
\textsc{Random-FullPool} & $59.5$ & $70.2$ & $56.1$ & $30.5$ & $73.8$ & $58.0$ \\
\textsc{No-Skill}        & $58.9$ & $69.1$ & $55.6$ & $29.6$ & $73.5$ & $57.3$ \\
\bottomrule
\end{tabular}
\end{table}

\paragraph{Result.} The hierarchy-then-LM \textsc{MetaBrowse} policy wins in every domain, averaging $68.9\%$ against $66.0\%$ for \textsc{BM25}, $64.2\%$ for \textsc{BM25+Embed}, $60.0\%$ for \textsc{Embed}, $58.0\%$ for \textsc{Random-FullPool}, and $57.3\%$ for \textsc{No-Skill}. Its largest margins over the strongest retrieval-only baseline are on Excel ($+5.0$ pp), PPT ($+3.8$), and Blender ($+2.3$), where task fit and complementarity are not captured by lexical or vector similarity alone.

\section{Conclusion}
\label{sec:conclusion}
\ourmethod distills multimodal human references into a structured, executable Skill Wiki shared by offline construction and controlled online gap filling. Across seven authoring domains and four backends, skill access improves artifact quality by $+11.9$ points over no-skill agents and beats two agentic-harness baselines in $26$ of $28$ main-aggregate cells. Our results show that distilling skills from human-created resources gives software agents reusable procedural knowledge that improves over both no-skill agents and strong agentic harnesses across diverse authoring domains.

\bibliographystyle{plainnat}
\bibliography{reference}

\newpage

\appendix

\section{Per-Domain Details}
\label{app:domains}

We summarize the implemented domain backends. Main results in Section~\ref{sec:experiments} cover Web, Excel, PPT, Blender, Reaper, CAD, and UE5; ablations report the five-domain core (Web, Excel, Reaper, PPT, Blender), with CAD and UE5 included in the main comparison and held out from per-condition ablations to keep the within-table compute budget bounded.

\paragraph{Slide design (PPT).} Slides are generated as SVG and then rendered to native PowerPoint (\texttt{.pptx}) shapes, on a fixed 16:9 canvas. Categories: layout, typography, palette, chart, icon, photo, shell, and motion. Active wiki size at evaluation start: 996. Render path: deck playback is rendered to video/frame artifacts for scoring; static-only runs use the same rendered slide frames as a contact sheet.

\paragraph{Spreadsheet authoring (Excel).} openpyxl with optional xlsxwriter for formatting. Categories: data ingest, formula pattern, conditional format, chart, dashboard layout, named range. Active wiki size: 632. Render path: LibreOffice export to per-sheet pages, composited into a worksheet contact sheet.

\paragraph{Web (HTML/CSS/JS).} Vanilla HTML5 + CSS3 + ES2020. Categories: layout, animation, typography, interaction, component, theme. Active wiki size: 941. Render path: Playwright full-page screenshot at a 1280-pixel viewport.

\paragraph{3D scenes (Blender).} bpy (Blender 4.1) headless. Categories: geometry, material, lighting, camera, animation, environment. Active wiki size: 661. Render path: the saved scene or final PNG is rendered to a single hero-shot frame.

\paragraph{Audio production (Reaper).} REAPER ReaScript over MCP. Categories: instrument, effect chain, automation, mix, arrangement. Active wiki size: 934. Render path: WAV bounce, transcoded to a bounded MP3 preview for the audio judge; spectrograms are retained only as diagnostics.

\paragraph{2D drafting (CAD).} ezdxf with FreeCAD as a fallback renderer. Categories: primitive, constraint, dimensioning, hatching, layer, layout. Library size: 312. Render path: FreeCAD viewport screenshot from two orthographic angles, composited.

\paragraph{Real-time 3D (UE5).} Unreal Engine 5 driven through a UE5-MCP bridge that exposes editor-side Python actions (actor spawn, transform, material assignment, level lighting, sequencer keyframe, asset import) as MCP tool calls. Categories: actor placement, lighting, materials and PBR, level blueprint, sequencer/cinematic, environment, post-process. Library size: $417$. Render path: editor viewport \texttt{HighResShot} of the saved level from a fixed camera, composited with a Lit-mode pass for scoring.

\paragraph{Skill-entry size and inference-time context budget.} Each skill entry on disk bundles agent-facing frontmatter (\texttt{skill\_name}, \texttt{category\_path}, \texttt{applicability}, \texttt{tags}, \texttt{modalities\_present}), a structured prose body (\texttt{text/overview.md}: mechanism, use conditions, inputs, expected effects), and the executable code body (\texttt{code/skill.py} or domain-equivalent). Visual thumbnails are referenced by path and resolved to images only when the agent explicitly requests the visual modality, so they do not enter the text-token budget. Token estimates use a conservative four-characters-per-token rule on the concatenation of frontmatter, prose, and code, computed across every entry in the offline wiki snapshot used for the main comparison. Per-domain mean / median tokens per entry are PPT $3{,}934$ / $4{,}108$, Excel $3{,}207$ / $2{,}612$, Web $5{,}726$ / $5{,}479$, Blender $4{,}248$ / $3{,}937$, and Reaper $4{,}170$ / $3{,}925$; aggregated across these domains the mean is $4{,}332$ tokens (median $4{,}133$, $25$th percentile $3{,}367$, $75$th percentile $5{,}049$). At inference time \textsc{MetaBrowse} forms a BM25 shortlist of $K{=}20$ candidates over the wiki taxonomy and exposes their compact frontmatter (typically $200$--$500$ tokens each) to the language-model selector, which picks $n{=}5$ entries to expose to the agent in full. The skill-related context per task is therefore $\sim 4$--$10$k tokens for the selection step and $\sim 22$k tokens for the five fully expanded entries; total $\sim 26$--$32$k tokens, well within the working window of every agent backend used in the main comparison and independent of the total library size since both $K$ and $n$ are fixed before the run.

\section{Benchmark Construction and Evaluation Protocol}
\label{app:benchmark-protocol}
\label{app:metrics}

\paragraph{Benchmark construction.} Each benchmark item is a task brief, not a demonstration. A brief specifies the domain, a stable \texttt{brief\_id}, a slug, the natural-language request, and the required output path and artifact type. It does not name skills, source URLs, or solution steps. We generate candidate briefs from domain-specific prompt templates with fixed seeds, then screen them manually for clarity, feasibility under the domain backend, and absence of overlap with the resource corpora used to build the skill wiki. For each domain, we retain a pool of $80$ screened briefs. The main comparison, scaling study, and online/offline study use a fixed matched $N{=}80$ subset. Ablations use separate fixed $N{=}40$ subsets sampled from the same $80$-brief pool because evaluating every ablation on all $80$ tasks would exceed the available run budget; all conditions within a given ablation share the same sampled brief IDs. The online/offline study additionally constructs $T_{\text{novel}}$ from capability regions that the offline pool fails to cover during preliminary screening, so it is a stress test for gap filling rather than a standard-distribution benchmark.

\paragraph{Run orchestration.} A run is one cell in the Cartesian product of domain, brief, and condition. All cells use the same agent loop and the same domain MCP server; conditions change only the skill access policy, wiki root, pool whitelist, source set, entry format, or harness baseline specified by the experiment. Unless a model ablation explicitly changes the backend, the agent uses GPT-5.4 with reasoning effort low. Each run follows the same four-stage loop: plan the task, browse or retrieve candidate skills when the condition permits it, execute through the domain apply tools, and verify by rendering the artifact. The experiment directories are self-contained: \texttt{experiments/<name>/briefs.json} stores briefs, \texttt{runs/<domain>/<slug>\_<condition>/} stores artifacts, \texttt{logs/} stores the agent trace, and \texttt{scores/} stores the judge JSON. Scaling pools and offline/online splits are fixed before launch and written to disk, so reruns reuse the same skill IDs rather than resampling the library.

\paragraph{Rendering.} Scoring always observes rendered artifacts rather than source files. Web outputs are rendered with Playwright full-page screenshots. Excel workbooks are exported through LibreOffice to per-sheet pages and stitched into a contact sheet. Blender outputs are scored from the final rendered PNG or a headless hero-shot render of the saved scene. PPT decks are rendered as presentation playback artifacts: when motion or transitions are part of the brief, we record video/frame evidence from the rendered slideshow; for static-only cells, the same pipeline reduces to rendered slide frames arranged as a contact sheet. Reaper projects are bounced to WAV, transcoded to a bounded mono MP3 preview, and sent as audio to the judge. CAD and UE5 follow the same artifact-first rule.

\paragraph{Scoring.} Non-audio artifacts are judged by GPT-5.4 with reasoning effort low. Reaper is judged by an audio-capable GPT-4o-series model on the bounced audio. Judges are blinded to the condition label and receive only the task brief, the rendered artifact, and the domain rubric. Each domain has five paper-facing axes: PPT uses layout quality, content density, theme coherence, typography hierarchy, and overall polish; Excel uses structure clarity, data density, chart quality, theme consistency, and overall polish; Web uses visual design, hierarchy clarity, section richness, modernness, and overall polish; Blender uses lighting, composition, materials realism, scene complexity, and overall aesthetic; Reaper uses low-end balance, transient clarity, spectrum balance, arrangement dynamics, and overall mix. The score file records an integer $0$--$10$ raw score and a one-sentence justification for each axis; all paper tables and figures report these scores after multiplying by $10$. The overall metric is the unweighted arithmetic mean of the five axes. Missing, unopenable, too-small, or silent artifacts receive $\text{overall}{=}0$ and are counted as failures; failures are kept in the denominator when computing condition means.

\paragraph{Aggregation and reproducibility.} For matched comparisons, we aggregate over the same brief IDs within each domain and condition. Reported deltas are therefore paired by construction. Every score JSON stores the artifact path, render path, rubric breakdown, overall score, and judge metadata, which allows later re-aggregation without re-running the agent.

\paragraph{Brief generation is wiki-blind.} Benchmark briefs are generated by a separate language-model call whose prompt receives only (i) the domain name, (ii) a hand-curated per-domain guidance string that enumerates types of artifacts in the wild, and (iii) a fixed structural template that specifies output format and the required output path. The brief generator does not receive any skill identifier, wiki listing, library size, taxonomy tree, per-skill metadata, or aggregated benchmark statistic from prior runs. The per-domain guidance strings were authored once before the main benchmark and were not updated in response to subsequent benchmark scores.

\paragraph{Resource collection queries are taxonomy-driven, not brief-driven.} The construction operator's resource queries are derived from per-domain category lists in the taxonomy that was hand-designed before benchmark briefs were generated. The query templates do not consume benchmark brief text, brief slugs, or any benchmark-side identifier; resource ingestion is a function of the taxonomy alone.

\paragraph{Acceptance thresholds are frozen before benchmark scoring.} The acceptance predicate $A_{\mathcal{D}}$'s thresholds (minimum schema fields, deduplication hash policy, executability gate, per-domain minimum-size constants) were tuned on a small dev-side resource sample drawn before main-benchmark briefs were generated and were frozen before the main comparison. Thresholds are not re-tuned in response to benchmark failure rates or per-cell scores.

\paragraph{No feedback loop from benchmark to library.} Per-cell failure rates, judge scores, and qualitative error analyses computed from benchmark runs are stored in \texttt{scores/} alongside the run artifacts; they are not piped back into the brief generator, the resource collector, the taxonomy, or the acceptance predicate. The library snapshot used for the main comparison is the same snapshot used for the scaling and online/offline studies, and was frozen before any benchmark cell was scored.

\section{Harness Baseline Configurations}
\label{app:harness}

This appendix documents the CLI versions and invocation contract under which \textsc{ClaudeCode-H} and \textsc{Codex-H} were run. Both harnesses are off-the-shelf agentic CLIs; we treat them as baselines rather than as custom-built systems, and intentionally retain their default planning and tool-use loops so that the comparison reflects the product a practitioner would actually deploy rather than a stripped variant.

\paragraph{Common protocol.} Both harnesses see the same brief, the same per-domain backend (SVG-rendered \texttt{.pptx}, openpyxl, Playwright, bpy/Blender, REAPER ReaScript, ezdxf/FreeCAD, UE5 Python), and the same output-path contract as \textsc{w Skills} and \textsc{w/o Skills}. Neither has access to the Skill Wiki or to any \ourmethod-specific prompt. Within each model--domain cell of Table~\ref{tab:main-final-v2}, brief IDs ($N{=}80$ matched per domain), agent backend, judge model and decoding settings (temperature $0$, judge reasoning effort low), output-path contract, and the per-domain failure-handling rule (overall${=}0$ on missing or unscorable artifact) are held fixed across all four systems, so within-cell deltas isolate the execution interface. Both harnesses run inside a per-brief working directory and produce artifacts there; live web browsing is not used as a path to ground-truth answers in either harness.

\paragraph{ClaudeCode-H.} Anthropic Claude Code CLI, version \texttt{2.1.x}. The CLI is invoked in non-interactive print mode with the brief on standard input and the per-brief working directory added to the tool-access scope. The underlying agent backend (GPT-5.5 / GPT-5.4 / GPT-5.4 Mini / GPT-5.4 Nano) is selected per row of the main comparison via the CLI's model-selection flag and routed through the same model deployments used for \textsc{w Skills} and \textsc{w/o Skills}. We use the harness-default tool surface (file read/write/edit, shell, grep/glob, web fetch) and explicitly do not load a project-level instructions file or any skill files, so the agent has no \ourmethod-specific priors. Permission prompts are bypassed so that runs are hands-off; reasoning effort is set to \emph{low}, matching \textsc{w Skills} and \textsc{w/o Skills}.

\paragraph{Codex-H.} OpenAI Codex CLI, version \texttt{0.129.x}. The CLI is invoked in non-interactive exec mode under a per-backend profile that pins the underlying model. Across all profiles we set the approval policy to never (no human-in-the-loop), the sandbox to workspace-write (the agent may write within the per-brief workspace), reasoning effort to \emph{low} (matching \textsc{w Skills} and \textsc{w/o Skills}), and web search to cached (browsing is restricted to the cached snapshot rather than live retrieval). The model providers point at the same model deployments used for \textsc{w Skills} and \textsc{w/o Skills}, so judges, decoding seeds, and model identifiers are matched. As with \textsc{ClaudeCode-H}, no skill library is mounted and no \ourmethod-specific prompt is injected.

\paragraph{Reproducibility.} The supplementary material includes the per-brief invocation contract for each harness, the per-condition run directories with raw stdout/stderr, agent traces, and produced artifacts, and the per-condition score JSONs that allow regeneration of every cell of Table~\ref{tab:main-final-v2} without re-invoking the agent or the judge.

\section{Construction Operator and Acceptance Predicate}
\label{app:operator-detail}

This section disaggregates the construction operator $f_\theta$ into its prompt-based and deterministic components, and lists what the acceptance predicate $A_\mathcal{D}$ actually checks.

\paragraph{$f_\theta$: prompt-based distillation with deterministic post-processing.} The construction operator is implemented as a vision-capable language-model call, not as a fine-tuned model, with no learned parameters specific to \ourmethod. Per resource $r$, the operator (i) issues a per-domain query template against the resource connector to retrieve the resource bytes (video frames, repository tree, article text, or reference artifact), (ii) extracts modality-specific evidence using deterministic preprocessors (key-frame sampling for video, AST-aware code region extraction for repositories, paragraph segmentation for articles, image preprocessing for artifacts), and (iii) calls the language model with a per-domain prompt template that elicits a structured JSON skill payload (\texttt{skill\_name}, \texttt{category\_path}, \texttt{applicability}, \texttt{tags}, \texttt{text\_body}, \texttt{code\_body}, \texttt{visual\_caption}, \texttt{provenance}). A deterministic post-processor normalizes whitespace, validates JSON shape, computes a stable skill identifier from a SHA1 hash of (\texttt{domain}, \texttt{source\_path}, \texttt{extracted\_node\_index}), and writes the entry to disk. The language-model role is therefore prompt-engineering plus structured output, not learned scoring.

\paragraph{$A_\mathcal{D}$: five deterministic gates.} The acceptance predicate is composed of five deterministic checks, each of which can independently reject a candidate skill. None is implemented as an LM-as-judge call.
\begin{itemize}
\item \textbf{Completeness.} A schema validator requires (a) all required frontmatter fields populated, (b) a non-empty \texttt{text\_body} of at least a minimum prose length, and (c) at least one populated content modality. Rejected candidates have null or shorter-than-threshold fields.
\item \textbf{Provenance.} A file-system check requires the resolved \texttt{source\_path} (or \texttt{video\_url}) to point to a retrievable resource recorded in the connector's manifest. Rejected candidates either omit provenance or point to a resource that is not in the manifest snapshot.
\item \textbf{Deduplication.} A SHA1 hash on (\texttt{domain}, \texttt{source\_path}, \texttt{extracted\_node\_index}) yields a stable skill identifier, allowing a single source to produce multiple distinct skills. A candidate sharing an identifier with an existing accepted skill is collapsed into the existing entry rather than added separately. Near-duplicate detection is implemented as exact identifier match plus a normalized-name string-equality check on the same source.
\item \textbf{Modality consistency.} A file-system check verifies that every modality declared in \texttt{modalities\_present} resolves to a non-empty file at the expected sub-path. Rejected candidates declare a modality that is not on disk.
\item \textbf{Structural executability.} For domains with executable skill bodies, a sandboxed test harness imports and runs the \texttt{code\_body} against minimal sample inputs (and, where applicable, runs render and overlap gates on the produced artifact). The gate sets a Boolean \texttt{exec\_ok} flag in the entry's metadata; entries with \texttt{exec\_ok=False} are retained in the wiki as reference-only and are filtered out by the agent's default verified-only mode. The executability check therefore validates that the code is structurally runnable and produces a non-trivial artifact, not that it solves any particular benchmark task.
\end{itemize}

\noindent The split is therefore prompt-based distillation (one LM call per resource) plus deterministic post-processing and gating (five rule-based checks). The library is not curated by an LM-as-judge loop, and the gates do not consult the benchmark.

\section{Retrieval and Composition Quality}
\label{app:retrieval-quality}

The main paper reports end-to-end overall scores rather than disaggregated retrieval metrics. This section describes what the run logs contain so that retrieval-quality analyses can be reproduced from the released artifacts, and explains why the end-to-end comparison already constrains retrieval and composition quality.

\paragraph{What is logged per run.} For every (\texttt{brief\_id}, \texttt{condition}) pair, the released run directory contains: (i) the BM25 shortlist of the top-$K{=}20$ candidates with their lexical scores; (ii) the language-model selector's transcript including the $n{=}5$ chosen skill identifiers and a brief justification per pick; (iii) the agent's execution trace recording which selected skills were actually invoked, in what order, and with what arguments; (iv) per-call tool invocation success and any apply-tool error returns; and (v) the final scored artifact and judge JSON. These logs permit downstream computation of precision-at-$K$ against any reference relevance label, of selection-call success rates, and of per-skill invocation success rates without re-running the agent.

\paragraph{Why end-to-end already constrains retrieval quality.} The matched-budget design of the selection-strategy ablation (Table~\ref{tab:selection-updated-aligned}) holds the library, agent, judge, decoding seed, and candidate budget fixed across selection strategies; the only varying factor is how candidate skills are ranked and presented. Under that design, a strategy that retrieves irrelevant skills can fail in two observable ways: the language-model selector can refuse to compose with them (it is permitted to pick zero skills), in which case the agent falls back to free-form code and the score regresses toward the no-skill baseline; or the agent can attempt to compose with them and produce an artifact that the rubric scores poorly. Both failure modes show up as a lower overall score in the corresponding row, which is what we observe for the retrieval-only baselines (\textsc{Embed} $60.0\%$, \textsc{BM25+Embed} $64.2\%$, \textsc{Random-FullPool} $58.0\%$) relative to \textsc{Ours} ($68.9\%$).

\paragraph{Composition behaviour.} A run is a single ordered application of $n{=}5$ selected skills (or fewer, if the language-model selector picked fewer); we do not perform multi-pass replanning or arbitration between conflicting skills, so the composition step is deterministic given the selected set. Composition conflicts therefore manifest as failures in the apply tool's structured not-applicable returns or as artifact-level scoring drops; both are recorded in the run directory and contribute to the per-cell overall score.

\section{Judge Reliability}
\label{app:judge}

We sampled 17 task--artifact pairs uniformly across domains and obtained five-axis scores from three human raters and from the corresponding automatic judge (GPT-5.4 for non-audio artifacts, audio-capable GPT-4o-series judging for Reaper). We report Spearman $\rho$ between the judge and the rater median, and the intraclass correlation coefficient ICC(2,1) treating raters and judge as exchangeable. On the overall axis, $\rho = 0.71$ and ICC $= 0.66$; design and detail axes track human medians most closely, while utility shows the highest disagreement (judge tends to reward apparent functionality even when interactivity is absent). Per-axis numbers are tabulated in the supplementary material. We re-ran the judge on the same 17 pairs and obtained $\rho = 0.83$ between runs, confirming that judge variance is substantially below judge--human variance.

\section{Per-Cell Paired Outcome Counts and Wilcoxon Tests}
\label{app:per-cell-stats}

Table~\ref{tab:per-cell-stats} reports paired outcome counts for the matched $N{=}80$ briefs, the matched delta, and the paired Wilcoxon signed-rank $p$-value. The outcome counts partition each cell into strict \textsc{w Skills} wins, ties, and strict \textsc{w/o Skills} wins.

\begin{table}[h]
\centering
\caption{\textbf{Per-cell paired outcome counts and paired Wilcoxon $p$-values} for the \textsc{w Skills} vs.\ \textsc{w/o Skills} delta on matched $N{=}80$ briefs per cell.}
\label{tab:per-cell-stats}
\small
\begin{tabular}{ll|rrr|rr}
\toprule
\textbf{Model} & \textbf{Domain}
& \textbf{\textsc{w Skills} Win}
& \textbf{Tie}
& \textbf{\textsc{w/o Skills} Win}
& \textbf{Delta}
& \textbf{Wilcoxon $p$} \\
\midrule
GPT-5.4 & Web     & $76$ & $3$  & $1$  & $+13.70$ & $4.18\!\times\!10^{-14}$ \\
GPT-5.4 & PPT     & $78$ & $2$  & $0$  & $+9.40$  & $1.86\!\times\!10^{-14}$ \\
GPT-5.4 & Excel   & $80$ & $0$  & $0$  & $+17.80$ & $1.21\!\times\!10^{-14}$ \\
GPT-5.4 & Blender & $79$ & $1$  & $0$  & $+14.60$ & $1.45\!\times\!10^{-14}$ \\
GPT-5.4 & Reaper  & $62$ & $10$ & $8$  & $+4.10$  & $2.05\!\times\!10^{-9}$  \\
GPT-5.4 & CAD     & $66$ & $9$  & $5$  & $+7.00$  & $1.14\!\times\!10^{-10}$ \\
GPT-5.4 & UE5     & $80$ & $0$  & $0$  & $+38.20$ & $1.21\!\times\!10^{-14}$ \\
\midrule
GPT-5.5 & Web     & $75$ & $4$  & $1$  & $+13.40$ & $6.27\!\times\!10^{-14}$ \\
GPT-5.5 & Excel   & $49$ & $16$ & $15$ & $+3.10$  & $3.15\!\times\!10^{-4}$  \\
\bottomrule
\end{tabular}
\end{table}

The \textsc{w Skills} delta over \textsc{w/o Skills} is significant in every reported cell at $p < 10^{-3}$, with $8$ of $9$ cells at $p < 10^{-8}$.

\section{Per-Axis Main-Comparison Details}
\label{app:perdim}

Table~\ref{tab:main-perdim} reports the per-axis breakdown of the GPT-5.4 main-comparison cell. Each domain is scored on its own five-axis rubric (full protocol in Appendix~\ref{app:metrics}); raw axis means are multiplied by $10$ and reported as percentages. The five axis values for each domain average to the corresponding GPT-5.4 overall score in Table~\ref{tab:main-final-v2}. The \textsc{w Skills} column shows the mean and, in parentheses, the matched percentage-point difference $\Delta$ relative to \textsc{w/o Skills} (\dpos{green} for positive). 

\begin{table}[h]
\centering
\caption{\textbf{Per-axis details for the GPT-5.4 main comparison (\%).} $\Delta$ is the matched difference of \textsc{w Skills} over \textsc{w/o Skills}.}
\label{tab:main-perdim}
\small
\begin{tabular}{ll|cc}
\toprule
\textbf{Domain} & \textbf{Axis} & \textbf{w Skills ($\Delta$)} & \textbf{w/o Skills} \\
\midrule
\multirow{5}{*}{PPT}
  & layout quality       & $68.6$ (\dpos{$+10.4$}) & $58.2$ \\
  & content density      & $61.3$ (\dpos{$+11.5$}) & $49.8$ \\
  & theme coherence      & $71.9$ (\dpos{$+9.2$})  & $62.7$ \\
  & typography hierarchy & $63.8$ (\dpos{$+12.3$}) & $51.5$ \\
  & overall polish       & $58.4$ (\dpos{$+3.6$})  & $54.8$ \\
\midrule
\multirow{5}{*}{Excel}
  & structure clarity    & $78.8$ (\dpos{$+18.5$}) & $60.3$ \\
  & data density         & $74.9$ (\dpos{$+15.8$}) & $59.1$ \\
  & chart quality        & $79.2$ (\dpos{$+22.4$}) & $56.8$ \\
  & theme consistency    & $72.1$ (\dpos{$+14.4$}) & $57.7$ \\
  & overall polish       & $77.0$ (\dpos{$+17.9$}) & $59.1$ \\
\midrule
\multirow{5}{*}{Web}
  & visual design        & $80.2$ (\dpos{$+4.9$})  & $75.3$ \\
  & hierarchy clarity    & $84.8$ (\dpos{$+14.0$}) & $70.8$ \\
  & section richness     & $86.4$ (\dpos{$+30.3$}) & $56.1$ \\
  & modernness           & $78.3$ (\dpos{$+4.7$})  & $73.6$ \\
  & overall polish       & $82.3$ (\dpos{$+14.6$}) & $67.7$ \\
\midrule
\multirow{5}{*}{Blender}
  & lighting             & $46.2$ (\dpos{$+14.9$}) & $31.3$ \\
  & composition          & $45.4$ (\dpos{$+15.6$}) & $29.8$ \\
  & materials realism    & $46.8$ (\dpos{$+18.1$}) & $28.7$ \\
  & scene complexity     & $39.7$ (\dpos{$+12.2$}) & $27.5$ \\
  & overall aesthetic    & $42.4$ (\dpos{$+12.2$}) & $30.2$ \\
\midrule
\multirow{5}{*}{UE5}
    & instruction correctness & $64.7$ (\dpos{$+24.5$}) & $40.2$ \\
    & scene design            & $71.4$ (\dpos{$+40.7$}) & $30.7$ \\
    & visual quality          & $60.2$ (\dpos{$+28.9$}) & $31.3$ \\
    & technical validity      & $68.4$ (\dpos{$+36.3$}) & $32.1$ \\
    & overall effect          & $71.8$ (\dpos{$+60.6$}) & $11.2$ \\
\midrule
\multirow{5}{*}{CAD}
    & drafting completeness    & $64.8$ (\dpos{$+8.6$}) & $56.2$ \\
    & dimensioning annotation  & $48.2$ (\dpos{$+6.4$}) & $41.8$ \\
    & layer/linework quality   & $56.4$ (\dpos{$+7.1$}) & $49.3$ \\
    & layout readability       & $57.2$ (\dpos{$+6.4$}) & $50.8$ \\
    & overall polish           & $51.9$ (\dpos{$+6.5$}) & $45.4$ \\
\midrule
\multirow{5}{*}{Reaper}
  & low-end balance      & $78.8$ (\dpos{$+4.6$}) & $74.2$ \\
  & transient clarity    & $75.9$ (\dpos{$+3.6$}) & $72.3$ \\
  & spectrum balance     & $77.3$ (\dpos{$+3.9$}) & $73.4$ \\
  & arrangement dynamics & $76.2$ (\dpos{$+4.4$}) & $71.8$ \\
  & overall mix          & $78.3$ (\dpos{$+4.0$}) & $74.3$ \\
\bottomrule
\end{tabular}
\end{table}

\section{Online Acquisition Details}
\label{app:online-details}

\paragraph{Trigger and budget.} Online acquisition is activated only when the offline wiki fails to return an adequate candidate set for the requested capability. In the offline/online study, the offline pool is fixed at launch and the online arm is allowed to add at most $100$ newly searched and distilled skills. This budget is held fixed across task sets so that the comparison measures coverage gain rather than unbounded resource access.

\paragraph{Validation.} Online candidates use the same construction predicate as offline candidates: the entry must have sufficient text evidence, traceable provenance, non-duplicate metadata, and a valid modality bundle. If executable code is present, it must pass the domain's smoke check or be marked as reference-only. Rejected candidates are not exposed to the agent.

\paragraph{Pool separation.} Online entries are stored in a separate pool for the duration of the evaluation split and are not folded back into the offline wiki used by the main comparison or ablations. This prevents online search from silently changing the default skill library.

\clearpage
\section{Case Studies}
\label{app:case}

We pair one \emph{success} and one \emph{failure} case for each of five representative domains (Web, PPT, Excel, Blender, Reaper) on the GPT-5.4 backbone. \emph{Success} cases are examples where the skill arm clearly improves on the no-skill arm; \emph{failure} cases are manually authored diagnostic boundary probes where skill use exposes a visible limitation, such as partial grounding, unresolved parameter binding, or overly literal composition. We do not show pure missing-output failures. Each figure shows the two arms side by side.

These qualitative examples are manually authored boundary probes and are not included in the matched benchmark aggregates. They should therefore not be interpreted as counterexamples to the paired benchmark results. Instead, they are used to make the method's limitations concrete: distilled skills improve average artifact quality, but they can still fail when a retrieved skill is only partially grounded, when symbolic bindings do not resolve, or when the agent composes a skill too literally.

\paragraph{Web success: cottagecore restaurant landing page.} Brief: build a farm-to-table restaurant page (\emph{cream, mossy green, floral motifs, rounded serifs; cottagecore digital}) with a creative testimonials section. The skill arm produces a complete, content-rich long-scroll page with consistent typography and a styled testimonials block; the no-skill arm produces a thin scaffold-like page that reads as an unfinished template.

\begin{figure}[H]
\centering
\webcasegraphic{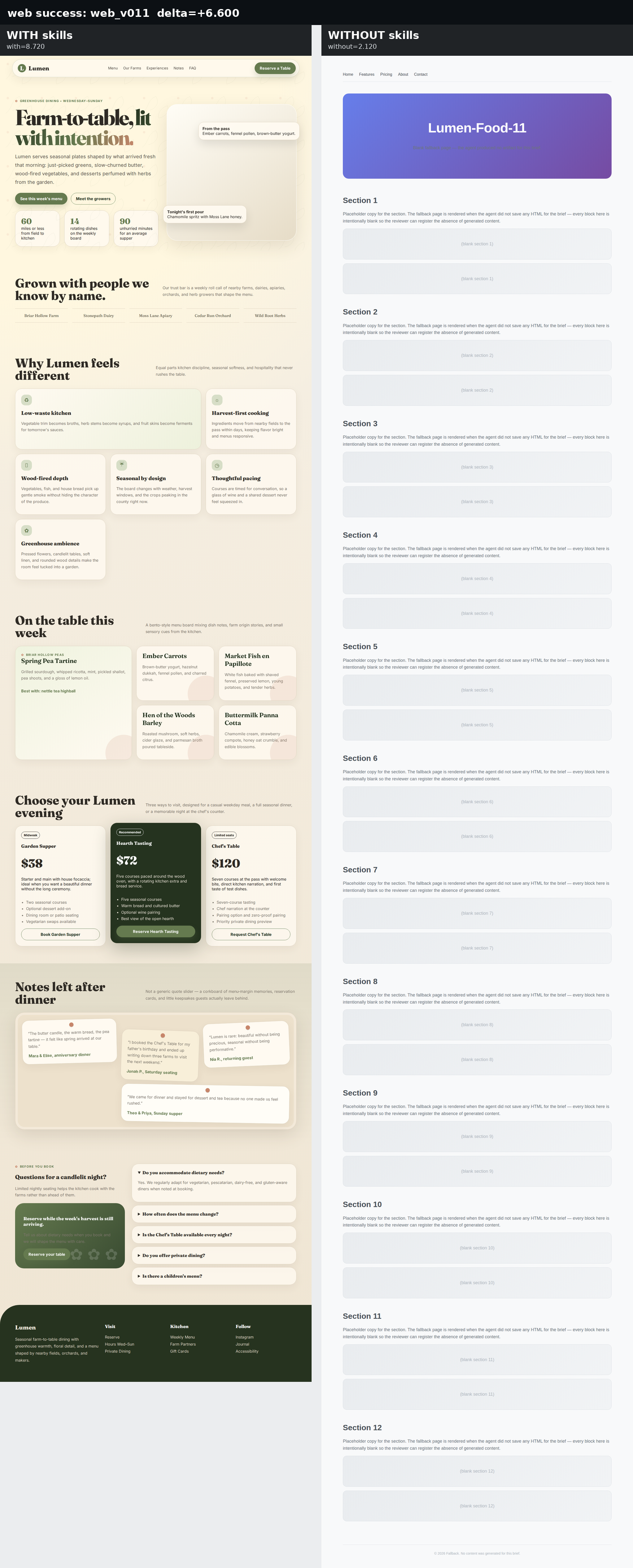}
\caption{\textbf{Web success.} \textsc{w Skills} (left) vs.\ \textsc{w/o Skills} (right).}
\end{figure}

\paragraph{Web failure: retro-futurist tabletop game page.} Brief: build a one-page site for a retro-futurist tabletop game. Both arms report \texttt{success: true} and \texttt{TASK\_COMPLETE}, so this is a quality failure rather than an execution failure: the skill arm is sparser and less finished, while the no-skill arm has fuller section coverage and stronger polish.

\begin{figure}[H]
\centering
\webcasegraphic{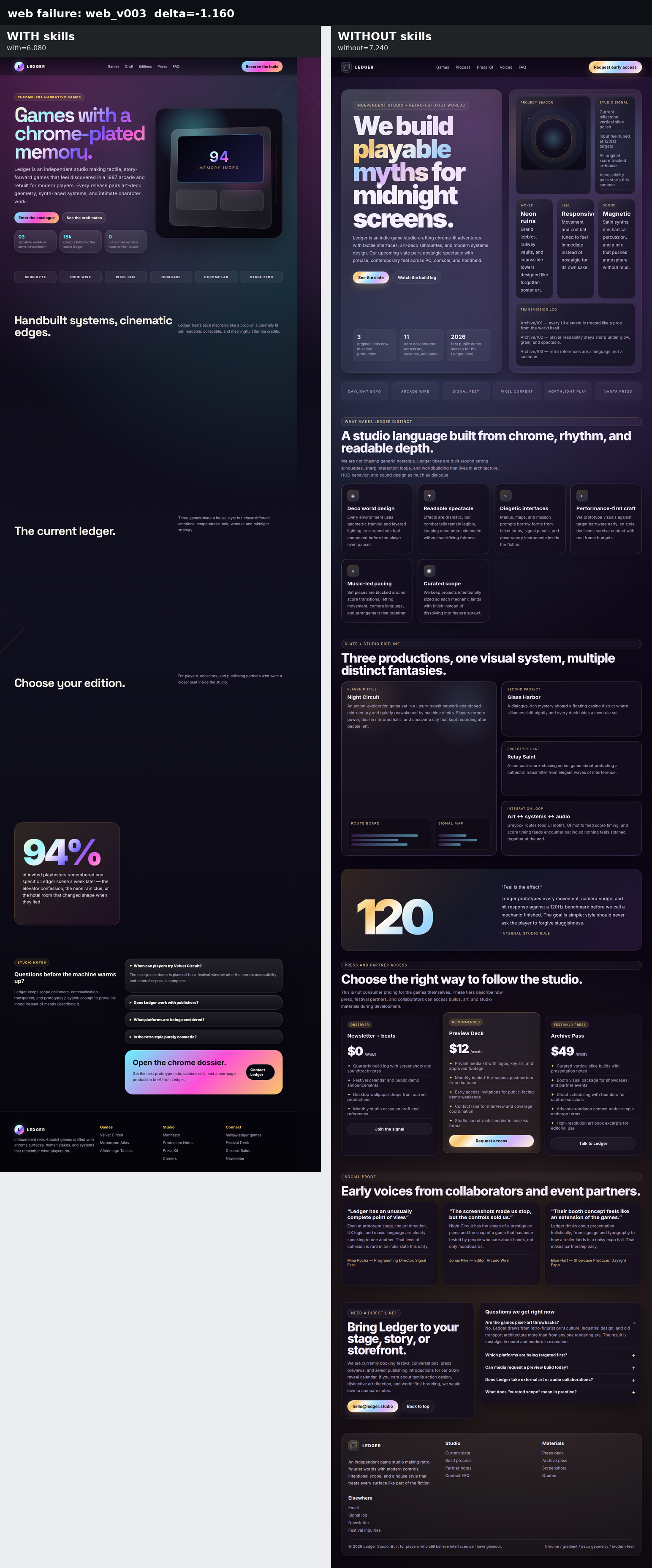}
\caption{\textbf{Web failure.} \textsc{w Skills} (left) vs.\ \textsc{w/o Skills} (right).}
\end{figure}

\paragraph{PPT success: corporate all-hands keynote.} Brief: build a $10$-slide internal all-hands deck for a customer-success team (\emph{sober and corporate, blue/gray restrained}) with cover, agenda, dividers, content slides, and closing. The skill arm produces a deck with shell variety, dense per-slide content, and consistent theming; the no-skill arm produces a thinner deck with several placeholder-feeling slides.

\begin{figure}[H]
\centering
\casegraphic{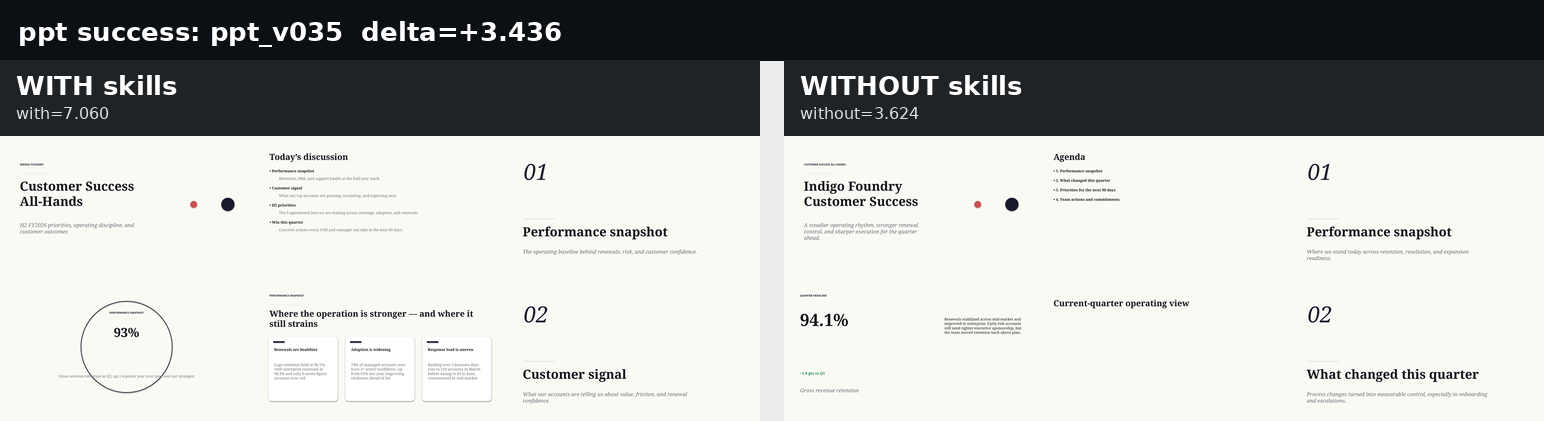}
\caption{\textbf{PPT success.} \textsc{w Skills} (left) vs.\ \textsc{w/o Skills} (right).}
\end{figure}

\paragraph{PPT failure: product strategy quarterly review.} Brief: build a $10$-slide quarterly review deck for potential acquirers (\emph{sober and corporate, blue/gray restrained}). The skill arm introduces unprocessed placeholder text and JSON-like fragments on the agenda slide and does not visibly inherit the chosen visual motifs; the no-skill arm is plainer but consistently styled.

\begin{figure}[H]
\centering
\casegraphic{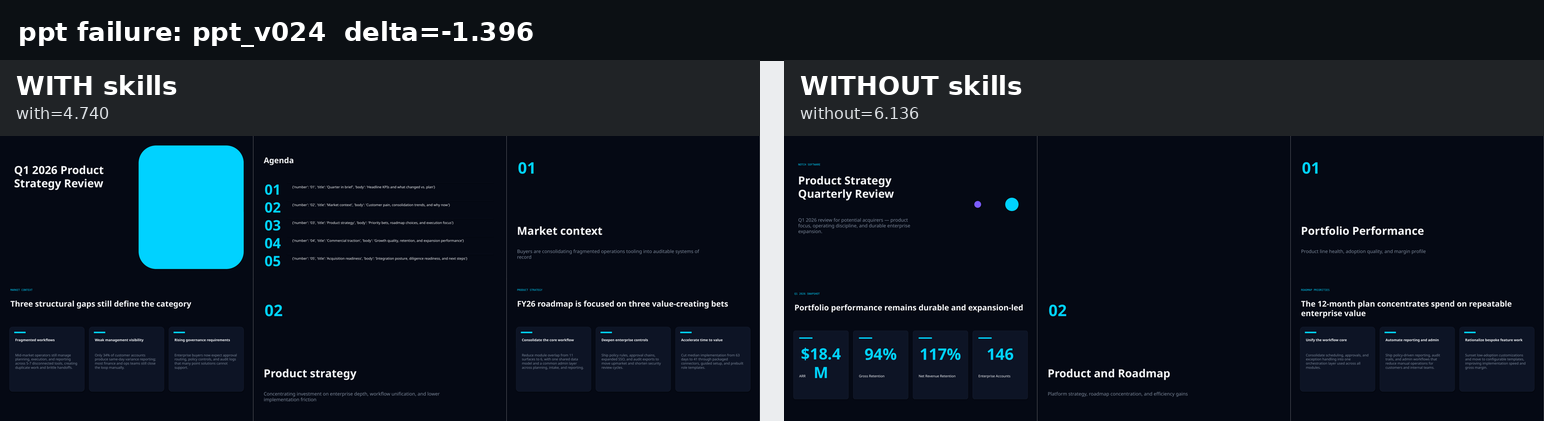}
\caption{\textbf{PPT failure.} \textsc{w Skills} (left) vs.\ \textsc{w/o Skills} (right).}
\end{figure}

\paragraph{Excel success: SaaS dashboard.} Brief: build a multi-sheet workbook with a summary dashboard, KPI blocks, and supporting detail sheets. The skill arm produces a summary-first workbook with KPI cards, dense data tables across sheets, and a coherent chart system; the no-skill arm produces a much sparser workbook with smaller, less polished tables and limited cross-sheet structure.

\begin{figure}[H]
\centering
\casegraphic{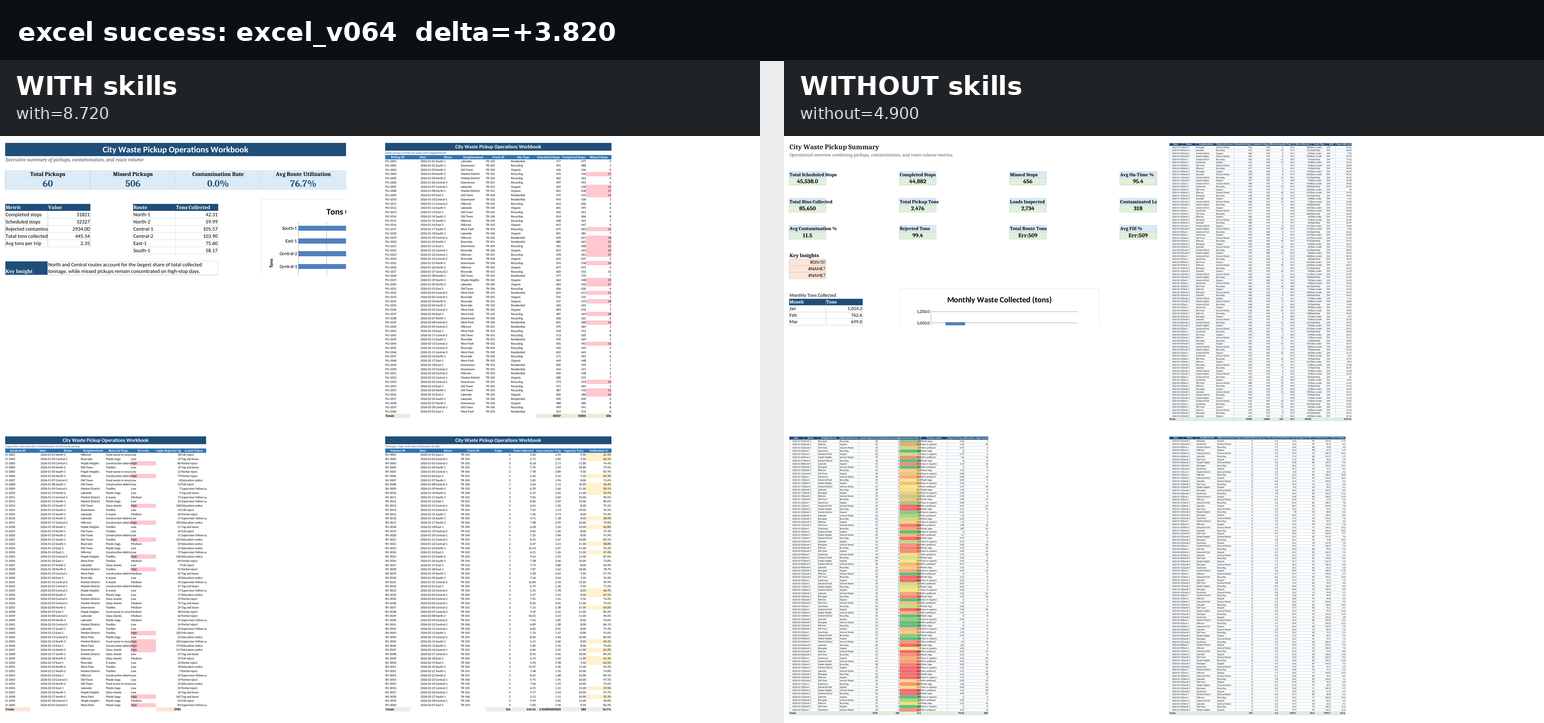}
\caption{\textbf{Excel success.} \textsc{w Skills} (left) vs.\ \textsc{w/o Skills} (right).}
\end{figure}

\paragraph{Excel failure: dashboard with broken formulas.} Brief: build an Excel workbook with dashboard and supporting detail sheets. The skill arm reuses dashboard components from the wiki but leaves visible \texttt{\#NAME?} formula errors on several cells; the no-skill arm avoids reusable components entirely and produces a cleaner, error-free aggregated dashboard. This case illustrates that a partially-grounded skill, with a referenced formula whose bindings did not resolve, can be worse than a simple bespoke layout.

\begin{figure}[H]
\centering
\casegraphic{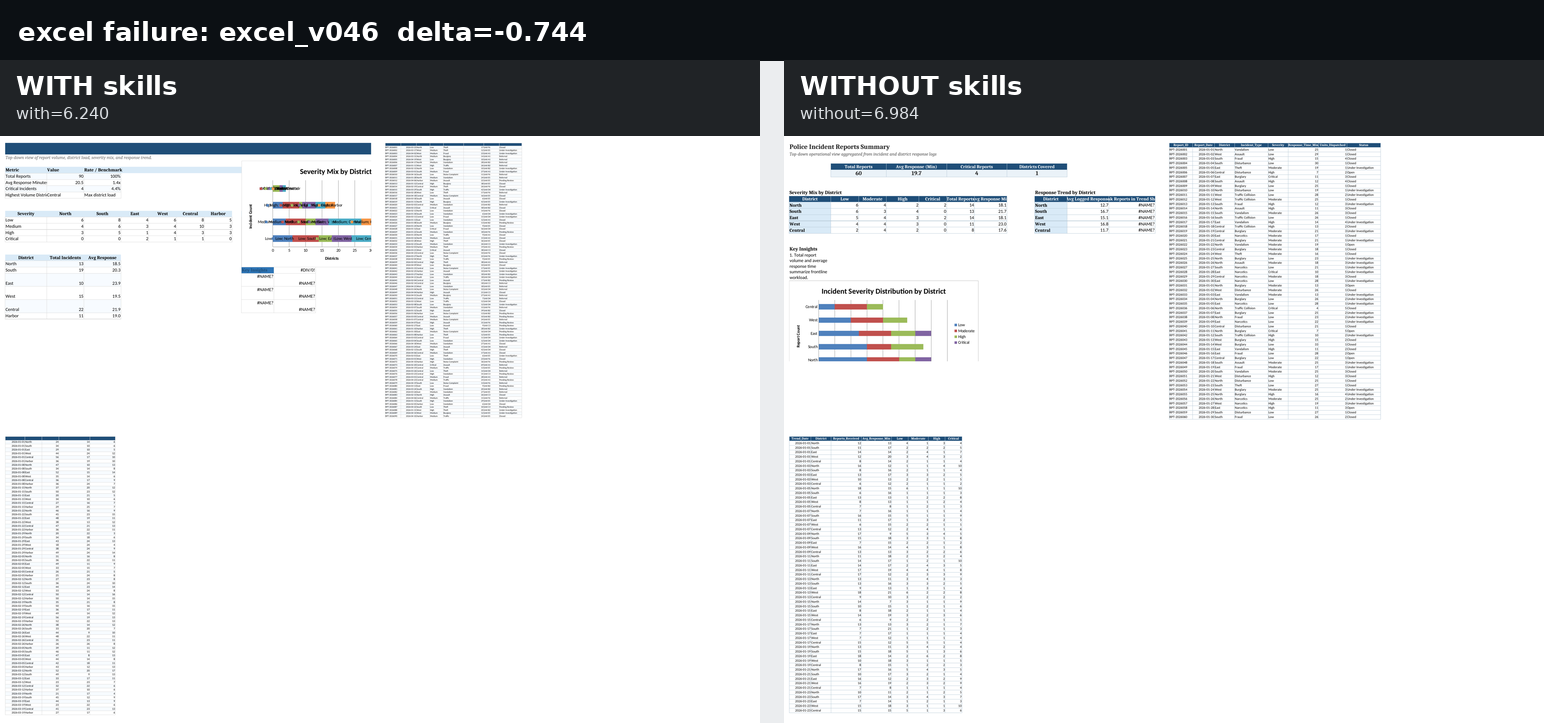}
\caption{\textbf{Excel failure.} \textsc{w Skills} (left) vs.\ \textsc{w/o Skills} (right).}
\end{figure}

\paragraph{Blender success: emerald jewelry hero shot.} Brief: render a single hero-shot 3D scene of an emerald ring with staged lighting and styled materials. The skill arm produces an identifiable hero subject with rim and key lighting (blue/amber treatment) and PBR materials; the no-skill arm produces a flatly lit primitive blockout that does not read as a jewelry render.

\begin{figure}[H]
\centering
\casegraphic{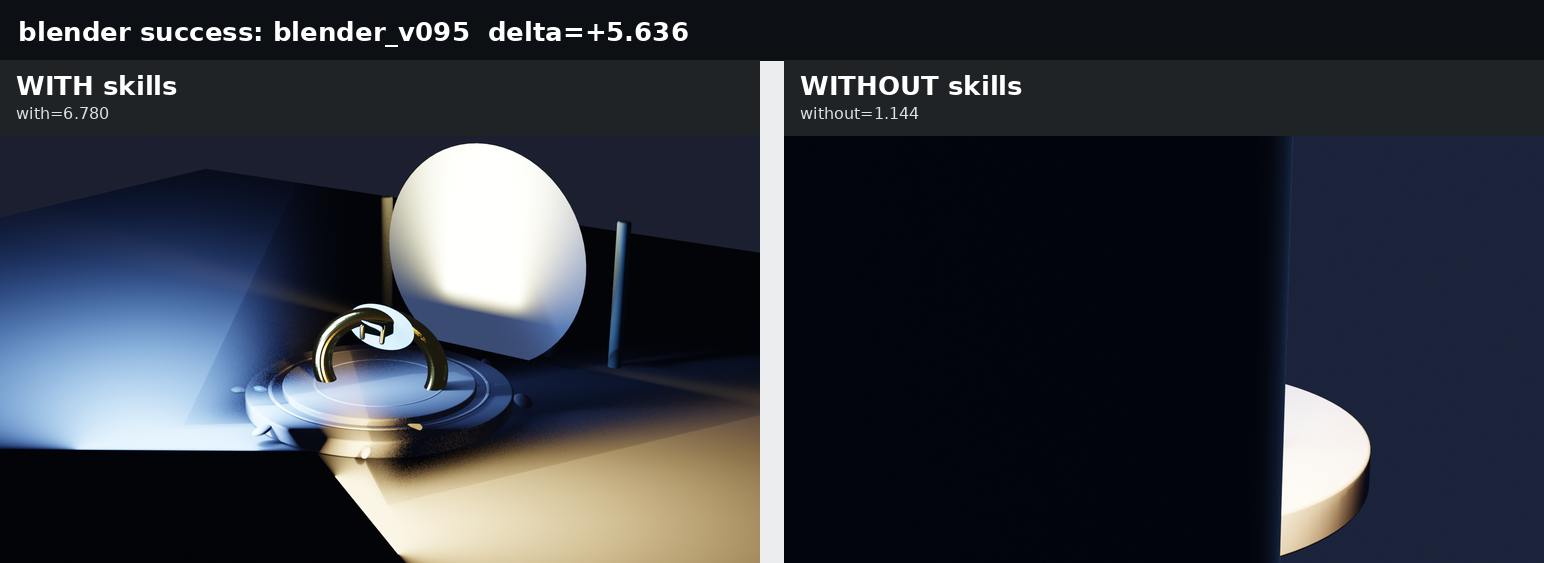}
\caption{\textbf{Blender success.} \textsc{w Skills} (left) vs.\ \textsc{w/o Skills} (right).}
\end{figure}

\paragraph{Blender failure: perfume bottle macro.} Brief: render a luxury macro of a glass perfume bottle. The skill arm inspects multiple glass and lighting skills but produces a washed-out final render that fails to show the bottle and its materials; the no-skill arm produces a simpler scene that at least reads as a recognizable bottle. This case illustrates that complex skill compositions can fail to ground when key parameters (background, exposure, camera) are not visibly inherited.

\begin{figure}[H]
\centering
\casegraphic{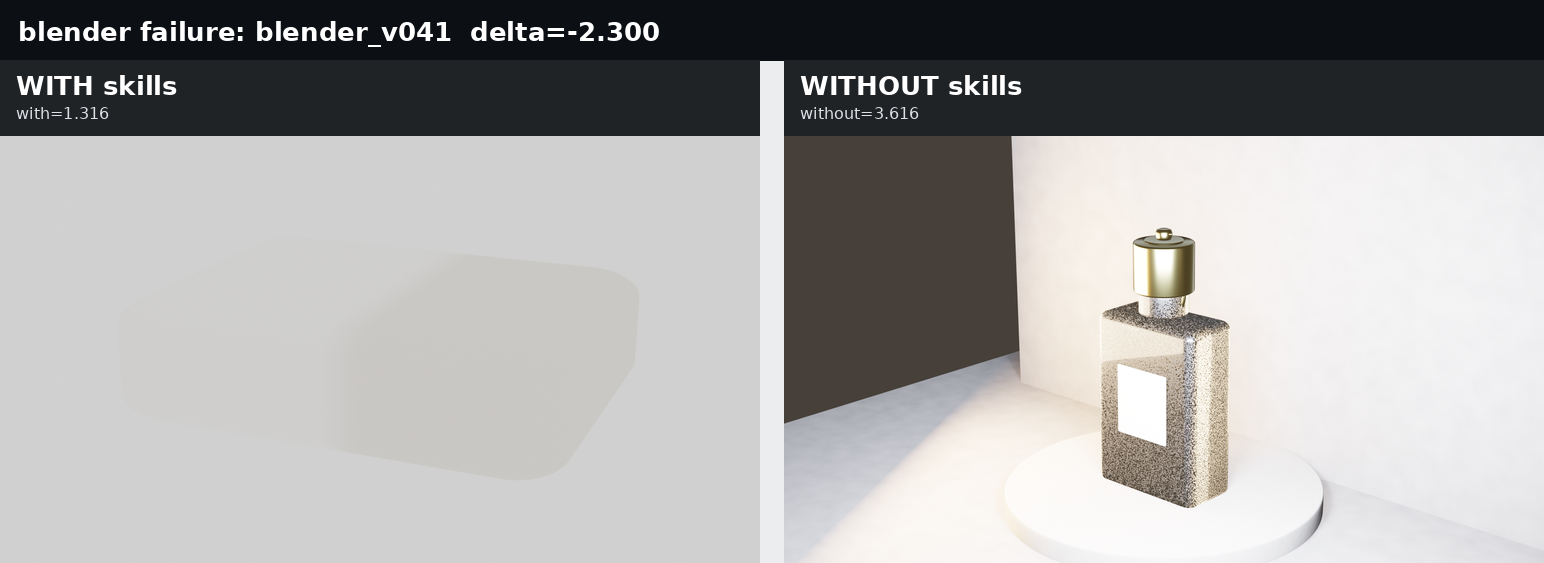}
\caption{\textbf{Blender failure.} \textsc{w Skills} (left) vs.\ \textsc{w/o Skills} (right).}
\end{figure}

\paragraph{Reaper success: deep house track.} Brief: produce a $30$--$60$\,s deep house track with sidechained bass, harmonic layering, and arrangement dynamics. The skill arm adds audibly grounded sidechain pumping, a distinct bassline, harmonic layering, and clearer intro/main/outro variation; the no-skill arm is competent but flatter in arrangement dynamics.

\begin{figure}[H]
\centering
\casegraphic{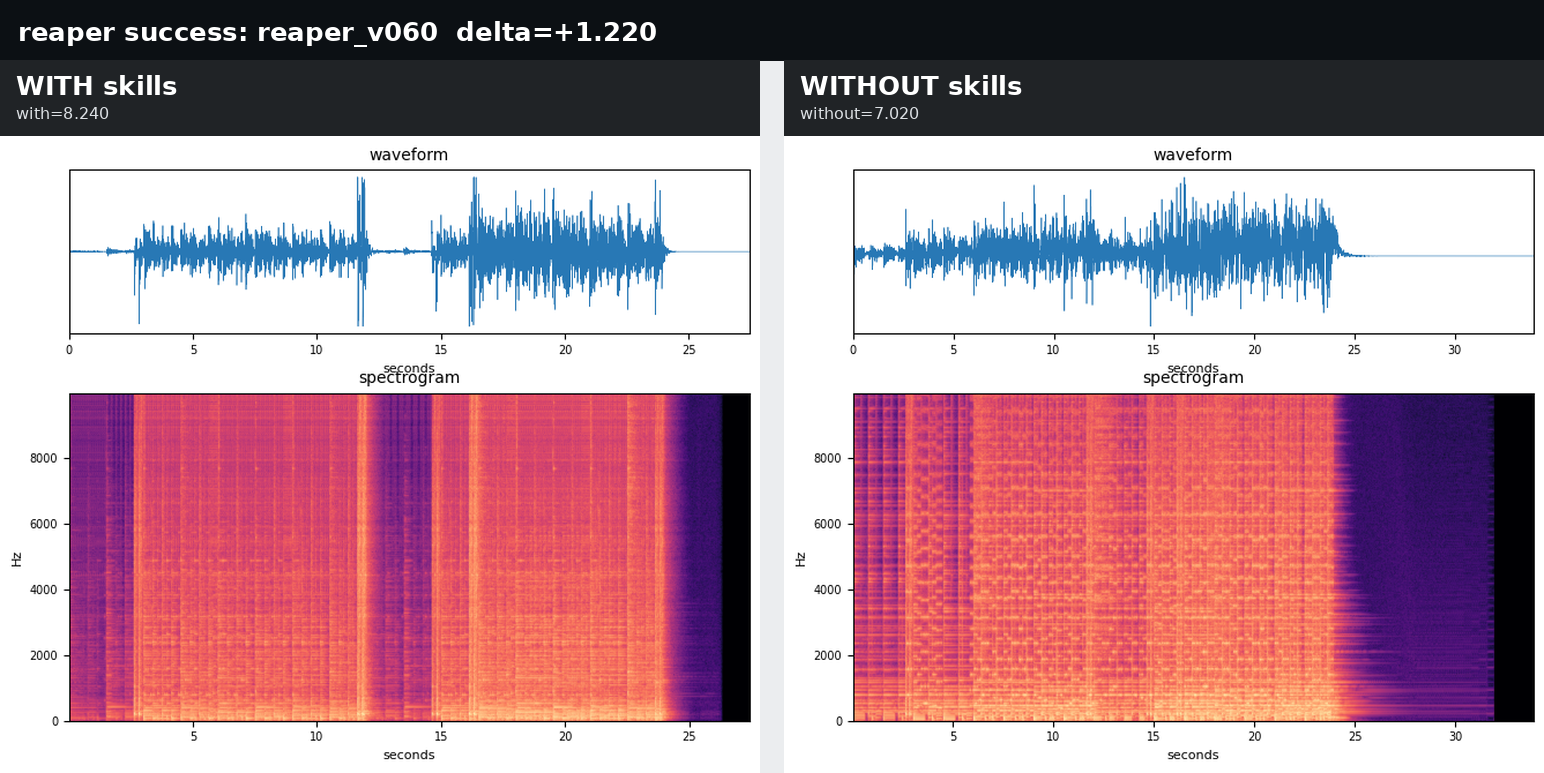}
\caption{\textbf{Reaper success.} Spectrogram and waveform of \textsc{w Skills} (top) vs.\ \textsc{w/o Skills} (bottom).}
\end{figure}

\paragraph{Reaper failure: less-dynamic arrangement.} Brief: produce a short genre track with intro/main/outro variation. The skill arm is competent but stays in one section throughout; the no-skill arm produces stronger intro/main/outro variation and overall mix polish. Reaper has the smallest no-skill-to-skill margin in the main aggregate; this case is consistent with that pattern.

\begin{figure}[H]
\centering
\casegraphic{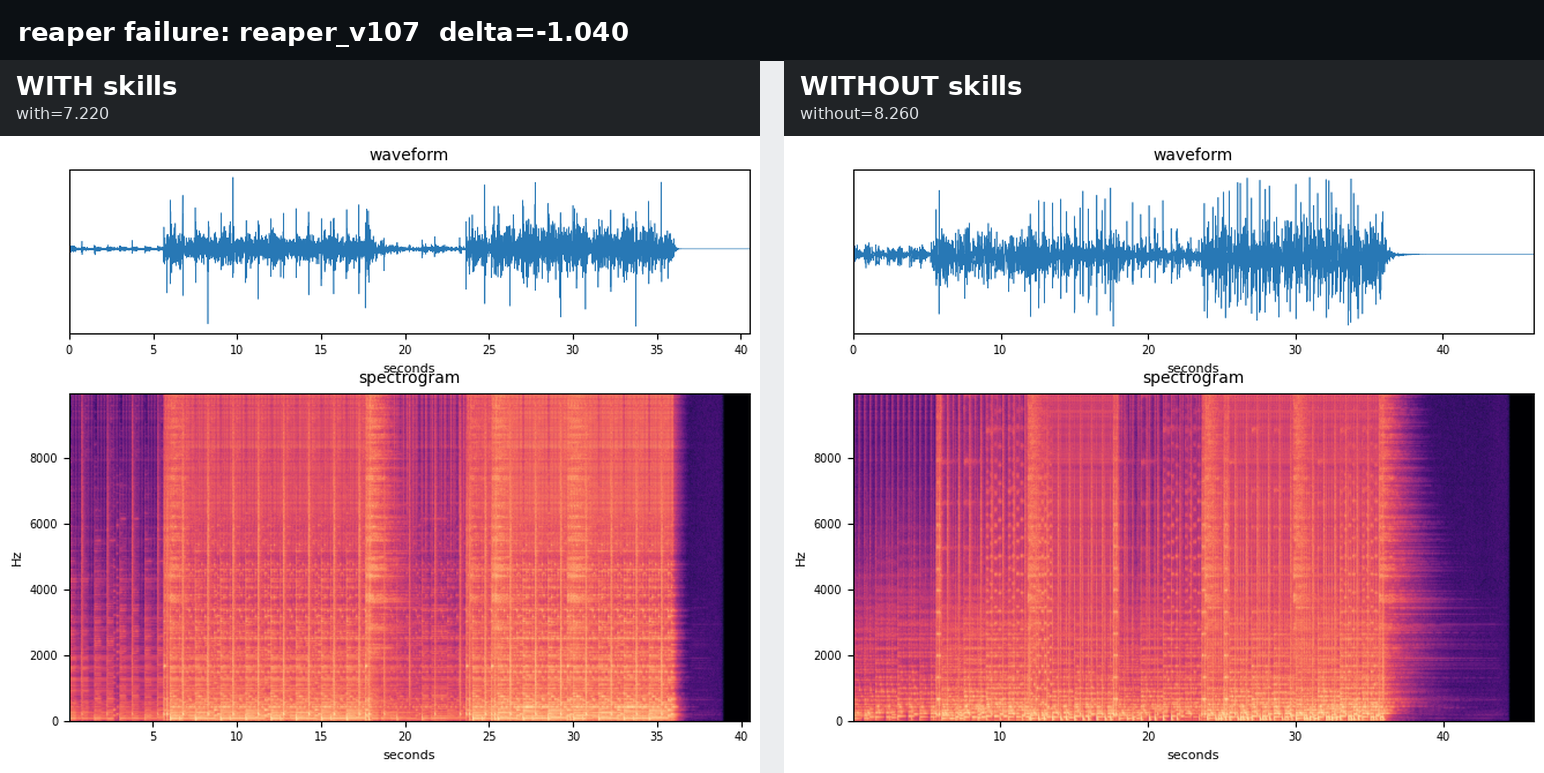}
\caption{\textbf{Reaper failure.} Spectrogram and waveform of \textsc{w Skills} (top) vs.\ \textsc{w/o Skills} (bottom).}
\end{figure}

\paragraph{What the failures reveal.} Across the five failure cases, two recurring patterns emerge: (i) \emph{partial grounding}, where a skill is selected and its surface pattern is borrowed but key bindings or parameters do not resolve in the final artifact (Web tabletop, Excel \texttt{\#NAME?}, Blender perfume); and (ii) \emph{conservative composition}, where the skill arm sticks closer to a single pattern and loses the variation a less-anchored agent would have introduced (PPT placeholder, Reaper less-dynamic). These patterns are consistent with the \textsc{Random-FullPool} and \textsc{Embed} retrieval baselines underperforming \textsc{Ours} in the selection ablation: a skill is only as useful as the agent's ability to bind its parameters, and stronger selection narrows this binding cost.

\clearpage

\section{User Study}
\label{app:userstudy}

We validate the main comparison with a blinded human A/B study. Artifact pairs are sampled from Section~\ref{sec:exp-main} balanced across the seven domains, and five human raters per pair view anonymized renderings side by side and choose the better artifact or declare a tie. We use this human study as an external validation of the automatic judge's preference direction, not as a training or selection signal.

Inter-rater agreement is Krippendorff's $\alpha = 0.58$. Table~\ref{tab:user-study} reports the per-domain breakdown as a micro-average over all $200$ individual ratings. \textsc{w Skills} wins $136$ of $200$ individual ratings ($68.0\%$), with $41$ ties ($20.5\%$) and $23$ \textsc{w/o Skills} wins ($11.5\%$); excluding ties, \textsc{w Skills}'s win rate is $85.5\%$. The smallest margin is on CAD ($77.8\%$ excl.\ ties), and the largest is on UE5 ($95.7\%$). The ordering of per-domain win rates broadly tracks the ordering of judge overall-score deltas, supporting the use of the automatic judge as the primary signal in the main paper.

\begin{table}[h]
\centering
\caption{\textbf{Per-domain human A/B study against \textsc{w/o Skills}.} Five raters per pair on anonymized side-by-side renderings. Aggregation is a micro-average over $200$ individual votes. \emph{Win rate excl.\ ties} excludes ties from the denominator.}
\label{tab:user-study}
\small
\resizebox{\textwidth}{!}{
\begin{tabular}{l|r|ccc|c}
\toprule
\textbf{Domain}
& \textbf{Total Votes}
& \textbf{\textsc{w Skills} Win (\%)}
& \textbf{Tie (\%)}
& \textbf{\textsc{w/o Skills} Win (\%)}
& \textbf{Win Rate excl.\ Ties (\%)} \\
\midrule
Excel   & $30$  & $23$ ($76.7$) & $4$ ($13.3$) & $3$ ($10.0$) & $88.5$ \\
Blender & $30$  & $20$ ($66.7$) & $7$ ($23.3$) & $3$ ($10.0$) & $87.0$ \\
Web     & $30$  & $19$ ($63.3$) & $7$ ($23.3$) & $4$ ($13.3$) & $82.6$ \\
PPT     & $30$  & $20$ ($66.7$) & $6$ ($20.0$) & $4$ ($13.3$) & $83.3$ \\
Reaper  & $30$  & $18$ ($60.0$) & $8$ ($26.7$) & $4$ ($13.3$) & $81.8$ \\
UE5     & $25$  & $22$ ($88.0$) & $2$ ($8.0$)  & $1$ ($4.0$)  & $95.7$ \\
CAD     & $25$  & $14$ ($56.0$) & $7$ ($28.0$) & $4$ ($16.0$) & $77.8$ \\
\midrule
\rowcolor{blue!10}
\textbf{Overall} & $\mathbf{200}$ & $\mathbf{136}$ ($\mathbf{68.0}$) & $\mathbf{41}$ ($\mathbf{20.5}$) & $\mathbf{23}$ ($\mathbf{11.5}$) & $\mathbf{85.5}$ \\
\bottomrule
\end{tabular}
}
\end{table}

\section{Skill Library Schema and Storage}
\label{app:schema}

Each skill is materialized as a modality bundle:

{\small
\begin{verbatim}
skills_wiki/<domain>/<skill_id>/
  source/        # provenance and resource references
  text/          # overview, mechanism, applicability, inputs
  visual/        # thumbnail, screenshot, render, or diagram
  code/          # executable or adaptable procedure, if available
  meta.json      # category path, tags, source type, validation status
\end{verbatim}
}

The metadata includes \texttt{skill\_id}, \texttt{skill\_name}, \texttt{domain}, \texttt{category\_path}, \texttt{tags}, \texttt{applicability}, \texttt{source.type}, provenance fields, and validation status. The prose body follows the same semantic sections used in the main text: mechanism, use conditions, inputs, and expected effects. This storage format is an implementation of the abstract tuple $s = (p, x_{\text{text}}, x_{\text{visual}}, x_{\text{code}}, m)$ from Section~\ref{sec:lm-wiki}.

\section{Limitations}
\label{app:limitations}

Most scores route through GPT-5.4 vision on rendered artifacts (Reaper through an audio-capable GPT-4o-series judge); judge--human agreement on the 17-task subsample (Appendix~\ref{app:judge}) is acceptable, and a blinded human A/B study with five raters per pair across all seven domains (Appendix~\ref{app:userstudy}) independently corroborates the preference direction reported in the main comparison. We do not claim generalization to domains lacking either a programmatic tool interface or a public stream of procedural content.

Online acquisition adds search, distillation, and validation latency on top of the normal benchmark pass. We deliberately keep it outside the main comparison and evaluate it only on a fixed online pool against $T_{\text{novel}}$, where gap-filling is the question being asked; folding online search into the default pipeline would conflate library-scaling effects with uncontrolled context expansion at test time. The online/offline study therefore measures coverage gain on capability regions known to be insufficient, not unbounded test-time recall.

Our retrieval-style baselines (\textsc{Embed}, \textsc{BM25+Embed}) operate over the distilled skill library, not over the raw resource corpus (transcripts, repository chunks, articles) under a matched token budget. We leave a same-budget raw-resource retrieval comparison to future work.


\clearpage

\end{document}

%% file: abstract.tex
\begin{abstract}
Skills are a useful abstraction for software agents, turning human and agent
experience into reusable procedural knowledge. Yet existing skill libraries are
mostly hand-written, text-centric, or derived from agent traces, leaving tutorial
videos and other multimodal human resources largely underused. We present
\ourmethod, a framework that distills multimodal resources---tutorial videos,
repositories, articles, and reference artifacts---into executable skills for
software agents. \ourmethod organizes these skills as a hierarchical multimodal
Skill Wiki, where each entry combines structured text, code, visual examples,
metadata, and provenance. This design preserves complementary signals from
different resources: videos capture temporal operations and visual effects, code
captures executable tool patterns, and articles or artifacts provide conceptual
and stylistic grounding. At inference time, agents retrieve and compose relevant
skills from the wiki; when coverage is insufficient, the same construction
operator can acquire new skills online. Across seven practical authoring domains,
\ourmethod improves average overall score by $+11.9$ percentage points over no-skill
agents and outperforms strong harness baselines in $26$ of $28$ main-aggregate
model--domain cells. Ablations confirm
the value of multimodal skill format, hierarchical organization, source diversity,
selection strategy, and online acquisition.
\end{abstract}

%% file: intro.tex
\section{Introduction}

Large language model agents~\citep{luo2025largelanguagemodelagent} are increasingly expected to do more than answer questions:
they must operate software, call tools, inspect intermediate results, and produce
high-quality artifacts such as slide decks, spreadsheets, web pages, 3D scenes,
CAD designs, and audio projects. In these settings, success often depends less on
isolated factual knowledge than on reusable procedural know-how: how to decompose
a goal, which tool or API pattern to use, what intermediate state to inspect, and
how to recover when an operation fails. We refer to such reusable procedural
knowledge as \emph{skills}. In the agent era, skills provide a natural abstraction
from \emph{experience} to \emph{reusable expertise}: they convert observations of
how tasks are solved into compact instructions, code fragments, visual references,
and execution recipes that can be repeatedly invoked by future agents.

Recent skill-based agent systems~\citep{xu2026agentskillslargelanguage} have already shown that such abstractions can be
highly valuable. Code agents and software assistants benefit from reusable tool-use
routines, while manually authored skill hubs have become a practical way to encode
expert conventions for document editing, programming, data analysis, and other
text-heavy workflows. However, existing skill libraries are still largely limited
by how they are created: they are either hand-written by experts, accumulated from
an agent's own interaction traces, or mined from text/code resources. This leaves
a major source of human expertise underused. For many commercial and creative
software tasks, the most natural way humans learn is not by reading static
documentation, but by watching tutorials, demonstrations, and screen-recorded
workflows. A video tutorial can reveal the temporal order of operations, the
visual effect of each editing step, and tacit design choices that are difficult to
express in text alone.

This observation motivates a broader question: \emph{Can we automatically distill
skills from multimodal human-created resources, especially tutorial videos, and
use them to build a scalable skill library for software agents?} The answer is not
straightforward. Although frontier models have seen massive amounts of text during
pretraining, and text resources are routinely exploited through retrieval and
search, high-dimensional multimodal resources remain much harder to use
effectively at inference time. Directly placing raw videos into an agent's memory
is expensive, redundant, and often impractical. A single tutorial may contain
minutes of irrelevant setup, repeated narration, and visual details that are
important only at a few key moments. At the same time, compressing video into a
plain text summary discards precisely the information that makes video useful:
dynamic operations, before-after visual changes, animation quality, spatial
layout, timing, and tool interaction order. A practical system must therefore
extract the procedural signal from videos and other resources, normalize it into
a reusable representation, and organize the resulting knowledge so that agents
can efficiently retrieve and execute the right skill for a new user request.

We introduce \ourmethod, a framework for distilling executable skills from
human-created resources and organizing them into a maintainable multimodal skill
library for software agents. Given multimodal resources---including tutorial
videos, source repositories, articles, documentation, and reference artifacts---
\ourmethod extracts domain-specific skills and stores them in a hierarchical
Skill Wiki. Each skill is represented as a multimodal entry that may contain
structured text, executable or adaptable code, visual examples, metadata, and
provenance. This design makes the skill library more than a flat collection of
retrieved passages: text explains applicability and mechanism, code provides
tool-grounded execution patterns, and visual examples preserve layout, style,
motion, and other perceptual information that text alone under-specifies.

A central design choice of \ourmethod is to treat skill construction and skill use
as a unified pipeline. Offline, we distill large-scale resources into a domain
wiki for important commercial software scenarios. At inference time, given a user
requirement, the agent first navigates the hierarchical index to form a candidate
skill pool, then reads the relevant multimodal entries and composes them during
execution. This hierarchical organization improves over flat retrieval because it
encodes domain structure, narrows the search space, and exposes skill candidates
at the appropriate level of abstraction. Moreover, when the offline library does
not cover a requested capability, the same resource-to-skill operator can be
invoked online to search for new resources, extract additional skills, and
incrementally extend the library. The resulting system is therefore not a fixed
prompt collection, but a growing and maintainable procedural memory.

We study \ourmethod across seven practical software-authoring domains, including
slide design, web page generation, spreadsheet authoring, Blender scene creation,
CAD design, UE5 scene construction, and music production. These domains are chosen
because they require different forms of procedural knowledge: some depend heavily
on code and API conventions, some on visual design and layout, and others on
temporal operations that are naturally conveyed by video. Our experiments show
that skill access consistently improves agent performance across model backends
and domains. Across our seven authoring benchmark suites (Web, Excel, Reaper,
PPT, Blender, CAD, UE5), \ourmethod improves the average overall
score by $+11.9$ percentage points over the same agents without skills, and
outperforms strong agentic-harness baselines in $26$ of $28$ main-aggregate cells.
Ablations further show that the hierarchical wiki interface, source diversity,
multimodal skill format, library scale, online skill acquisition, and
hierarchy-then-LM selection strategy each contribute to the final performance.

Our contributions are summarized as follows:
\begin{itemize}
    \item \textbf{Resource-to-skill learning for software agents.}
    We formulate the problem of distilling reusable executable skills from
    multimodal human-created resources, with tutorial videos as a key
    underused source of procedural and perceptual knowledge.

    \item \textbf{A hierarchical multimodal skill library.}
    We propose a Wiki-based organization in which each skill combines structured
    text, executable or adaptable code, visual examples, metadata, and provenance,
    enabling agents to retrieve, inspect, and compose skills across diverse
    software domains.

    \item \textbf{A unified offline-online construction pipeline.}
    The same resource-to-skill operator is used both to build large offline skill
    libraries and to acquire new skills online when user requirements expose
    capability gaps, making the library maintainable and incrementally extensible.

    \item \textbf{A broad empirical study across commercial authoring tasks.}
    We evaluate \ourmethod on seven authoring benchmark suites. Skill access yields
    consistent gains over
    no-skill agents and strong harness baselines, while controlled ablations
    isolate the value of source mix, multimodal format, wiki organization,
    selection strategy, and online acquisition.
\end{itemize}